\documentclass[11pt,a4paper]{article}
\usepackage{amsmath,amssymb}
\usepackage{float}
\usepackage{dcolumn}
\newcommand{\diffd}{\mathrm{d}}
\newcommand{\alphas}{\alpha_{\text{s}}}
\newcommand{\alphaem}{\alpha_{\text{em}}}
\newcommand{\encee}{N_{\text{c}}}
\newcommand{\CF}{C_{\text{F}}}
\newcommand{\CA}{C_{\text{A}}}
\newcommand{\TR}{T_{\text{R}}}
\newcommand{\gs}{g_{\text{s}}}
\newcommand{\pt}{p_{\text{T}}}
\newcommand{\kt}{k_{\text{T}}}
\newcommand{\ptsq}{p_{\text{T}}^{2}}
\newcommand{\ptsqmin}{p_{\text{T,min}}^{2}}
\newcommand{\nf}{n_{\text{f}}}
\newcommand{\dotqcdqedew}{\text{Dot$\otimes$QCD$\oplus$QED$\oplus$EW}}
\newcommand{\qcdqedew}{\text{QCD$\oplus$QED$\oplus$EW}}
\newcommand{\qcdqed}{\text{QCD$\oplus$QED}}
\newcommand{\qcdonly}{\text{QCD only}}
\usepackage{jheppub}
\makeatletter\gdef\@fpheader{}\makeatother

\title{Testing and Tuning the Interleaved QCD{+}QED{+}EW Angular-Ordered Shower in Herwig~7}

\author[a]{M. R. Masouminia}
\author[b,c]{M. H. Seymour}
\author[b]{S. Sule}

\affiliation[a]{Institute for Particle Physics Phenomenology, Durham University, Durham, U.K.}
\affiliation[b]{Department of Physics and Astronomy, University of Manchester, Manchester, U.K.}
\affiliation[c]{Department of Theoretical Physics, CERN, Geneva, Switzerland}

\emailAdd{michael.seymour@manchester.ac.uk}
\emailAdd{siddharth.sule@manchester.ac.uk}

\abstract{
We investigate the predictions of the angular-ordered parton shower in Herwig~7, focusing on the interplay among interleaved QCD, QED, and electroweak emissions across three kinematic reconstruction schemes. The differences between the interleaving and reconstruction schemes are probed using LEP, Tevatron and LHC observables sensitive to soft wide-angle radiation, including event shapes, jet masses, jet widths, radial energy profiles and leading-jet angularities. Using comparisons to data and projections at 100~TeV, along with bin-wise envelope diagnostics to quantify the spread, we find that interleaving non-QCD sectors on the angular-ordered clock induces only small variation in the observables considered here. We then tune the interleaved shower and cluster hadronization parameters and see improvements in LEP charged-particle spectra, multiplicities, and LHC jet-substructure observables without degrading IRC-safe observables such as event shapes and jet rates. We conclude that, for the observables studied here, interleaving the three sectors preserves the tested QCD structure, while quantitative performance hinges on the reconstruction scheme and the details of hadronization.
}

\begin{document}
\noindent{\hfill \small CERN-TH-2026-171}

\noindent{\hfill \small MCnet-26-38}

\noindent{\hfill \small IPPP-26-29\\[0.1in]}
\maketitle
\flushbottom


\section{Introduction}
\label{sec:into}

Angular-ordered (AO) parton showers offer a principled approach for modelling colour coherence by enforcing an emission hierarchy in angle. In Herwig~7~\cite{Bahr:2008pv, Bellm:2015jjp, Bewick:2023tfi, Bellm:2025pcw}, this is systematised via three AO-compatible kinematic reconstruction schemes, the original $\pt$-preserving~\cite{Gieseke:2003rz, Bahr:2008pv}, virtuality-preserving from version 7.0~\cite{Bellm:2015jjp}, and current default dot product+veto~\cite{Marchesini:1983bm, Marchesini:1987cf, Bewick:2019rbu, Bewick:2021nhc} schemes, which control departures from the strict soft-collinear limit while preserving the soft-angle mapping. In addition, a process-independent electroweak (EW) module with helicity-resolved $W/Z/H$ emissions interleaved on the common AO clock has been introduced~\cite{Masouminia:2021kne, Masouminia:2026udz}. Delivered within the Herwig~7 line, these ingredients furnish a unified, interleaved $\qcdqedew$ AO shower. In this paper, we run tests to ensure that this model preserves the global QCD structure.

Since the subsequent analyses rely critically on preserving QCD structure under these interleavings, we briefly recall the essential physics and its manifestation in AO showers.\footnote{The recent availability of generalised AO splittings for BSM studies~\cite{Lee:2023hef, Kim:2026mfk} and an NRQCD-based quarkonium shower~\cite{Masouminia:2025kec} further motivates careful bookkeeping. However, these developments lie outside the scope of our present study.} The structure of multiple emissions in QCD requires a collinear bundle of partons to radiate at larger angles as a single source carrying its total colour charge. Consequently, additional small-angle branchings do not increase the large-angle radiation rate, and emissions that are well separated in angle behave independently. In AO showers, this structure is implemented by arranging successive branchings at decreasing angles, thereby accounting for the interference between unresolved emitters~\cite{Marchesini:1987cf, Gieseke:2003rz, Ellis:1991qj, Dokshitzer:1991wu}. Formally, the emission current for a soft gluon of momentum $k$ radiated from a pair of partons $i$ and $j$ with momenta $p_i$ and $p_j$ reads
\begin{equation}
\mathcal{M}^\mu(k) \;\sim\; \gs \left( {T^a_i \frac{p_i^\mu}{p_i \cdot k} + T^a_j \frac{p_{j}^\mu}{p_j \cdot k}} \right)\, .
\end{equation}
At small angles and in the large-$\encee$ limit, which neglects terms of relative order $1/\encee^2$, only emission from pairs of colour-connected partons contributes, which, upon squaring, yields the dipole pattern~\cite{Ellis:1991qj}
\begin{equation}
|\mathcal{M}(k)|^2 \;\propto\; -
\frac{\CA}2 \left( \frac{p_i}{p_i \cdot k} - \frac{p_j}{p_j \cdot k} \right)^2\, .
\end{equation}
At angles larger than the dipole opening angle, the gluon cannot resolve the difference between $p_i$ and $p_j$, and the current is approximately
\begin{equation}
\mathcal{M}^\mu(k) \;\sim\; \gs \left(T^a_i+T^a_j\right)
\frac{p_{ij}^\mu}{p_{ij} \cdot k}\, ,
\end{equation}
where $p_{ij}$ is an on-shell momentum approximately equal to $p_i+p_j$. That is, the gluon can only resolve
the net colour charge rather than the individual partons. This coherence mechanism underpins the Modified Leading Log Approximation.
In the massless, large-$\encee$ limit, and for a gluon-dominated cascade, it gives the schematic scaling
\begin{equation}
\langle n(Q) \rangle \sim \exp\!\left[
    \sqrt{\frac{16\encee}{\beta_0}\ln\frac{Q}{\Lambda}}
\right]\, ,
\end{equation}
with $\beta_0 = 11 - 2\nf/3$. Collinear effects can be incorporated through the appropriate splitting functions, with their corresponding colour factors, while heavy-quark masses can be treated through the quasi-collinear approximation or, more simply, through the dead-cone approximation. This scaling nevertheless captures the observed growth of jet multiplicities and supports the AO shower picture~\cite{Gieseke:2003rz, Dokshitzer:1991wu}.

In all configurations studied here, QCD emissions are implemented through the angular-ordering condition~\cite{Marchesini:1987cf, Gieseke:2003rz}. We therefore ask whether QCD predictions for observables are disrupted by interleaving QED and EW emissions. Interleaving places the different emission types on a common evolution clock while retaining the underlying QCD angular-ordering condition~\cite{Masouminia:2021kne, Bellm:2019zci}. The extension to generalised EW branchings, and possibly to more exotic BSM splittings, further amplifies the need to maintain this global structure. In Herwig~7, this is implemented using the three kinematic reconstruction schemes above. The angular-ordering constraint is the central ingredient that implements colour coherence, reproducing the LL soft-collinear structure and incorporating important NLL coherent effects by restricting subsequent emissions to smaller angles. The three schemes are therefore expected to agree in the strict soft-collinear limit. However, they are not logarithmically identical away from that limit: their mappings between $\tilde q$ and angle, their recoil prescriptions, and their treatment of dead zones, heavy-flavour thresholds and hard non-collinear branchings can affect NLL-level terms~\cite{Reichelt:2017hts, Bewick:2019rbu, Bewick:2021nhc}. Thus, while AO secures the theoretical basis of colour coherence, its quantitative manifestation in jet shapes, multiplicities, and other observables depends on the interplay between interleaving across the QCD, QED, and EW sectors and the chosen kinematic reconstruction scheme, motivating dedicated validation and tuning~\cite{Bellm:2019zci, Bewick:2023tfi}. This motivates a quantitative envelope comparison, which we will introduce in section~\ref{subsec:coh-summary}, in which interleaving-induced and scheme-induced variations are explicitly separated.

Including EW radiation in an AO QCD shower raises a structural issue already visible in the simpler case of mixing QCD emissions with photon radiation: once non-QCD emissions are interspersed with gluons, the exact soft-interference constraints are not, in general, representable as a strictly Markovian evolution in a single local evolution variable, since the admissible angular range for a QCD branching can depend on the colour-antenna history of the recoiling system~\cite{Seymour:1991xa}. A pragmatic resolution is to enforce QCD coherence by ordering QCD emissions in angle with respect to the most recent QCD branching, treating intervening non-QCD emissions as spectators to the QCD ordering (an approximation argued to be valid at least after azimuthal averaging)~\cite{Seymour:1991xa}. It was also noted that evolution variables tailored to strict angular ordering can misrepresent the true kinematic boundary when mixed QCD-EW histories permit strongly disordered angles, generating phase-space dead regions; this motivates treating angular ordering as a master clock for coherence while recognising that quantitative behaviour away from the strict soft-collinear limit depends on the detailed evolution-kinematics mapping and recoil choices. In parallel with developments in angular-ordered showers, electroweak radiation has been studied from several complementary directions, including electroweak Sudakov logarithms and their resummation~\cite{Denner:2000jv, Denner:2001gw, Chiu:2007yn, Chiu:2007dg, Chiesa:2013yma, Cuomo:2019siu}, electroweak splitting functions~\cite{Chen:2016wkt, Krauss:2014yaa, Nardi:2024tce, Dittmaier:2025htf}, electroweak parton-shower implementations~\cite{Christiansen:2014kba, Kleiss:2020rcg, Brooks:2021kji, Masouminia:2021kne, Masouminia:2026udz}, and NLO EW/QED matching to parton showers~\cite{Richardson:2010gz, Flower:2026byh, Flower:2026lui}.

At leading power, AO directly realises colour coherence: enforcing $\theta_{n+1}<\theta_n$ reproduces the wide-angle soft-gluon interference pattern and provides the LL structure together with the NLL coherent corrections associated with angular ordering. Full NLL accuracy for a given observable, however, also depends on the recoil prescription, phase-space boundaries, coupling scheme and the precise interpretation of the AO evolution variable~\cite{Bewick:2019rbu, Dasgupta:2018nvj}. Showers ordered in other variables (e.g.~$\pt$ or dipole/antenna $\kt$) can preserve the same pattern provided kinematics and recoil ensure an emission hierarchy monotonic with angle in the soft limit (global recoil, colour-dipole radiation, vetoed non-monotonic branchings)~\cite{Herren:2022jej, Preuss:2024vyu, vanBeekveld:2024wws, Helliwell:2025krf, Holguin:2026trd}. For non-QCD sectors, the constraint is looser: QED coherence is Abelian, so AO is not required for correctness but must not distort the gluon pattern when QCD and QED emissions compete; EW emissions are tempered by masses, yet at high scales their Sudakov/interference structure benefits from an ordering that does not upset the global QCD angular hierarchy. The kinematic reconstruction scheme affects the radiation pattern beyond the strict soft-collinear limit through the evolution-angle/formation-time mapping and recoil/dead-zone boundaries; in Herwig, the three kinematic reconstruction schemes share the coherent AO limit but differ for hard or massive branchings, and hence can differ in their logarithmic accuracy~\cite{Reichelt:2017hts, Bewick:2019rbu, Bewick:2021nhc}. These differences can shift wide-angle radiation, heavy-flavour dead cones, and the competition with QED/EW splittings. Consequently, AO fixes the principle of coherence, while in practice predictions of jet shapes, multiplicities, and substructure depend on the evolution variable, recoil strategy, and interleaving prescription, and thus require dedicated validation and tuning~\cite{Bellm:2019zci, Bewick:2023tfi}.

For interactions beyond QCD, AO is sufficient, provided it preserves each sector's soft/collinear hierarchy and does not distort the gluon pattern when interleaved. In QED, the soft current is Abelian,
\begin{equation}
    J^\mu_{\text{QED}}(k)=e\sum_i Q_i\,\frac{p_i^\mu}{p_i\!\cdot\!k}\, ,
\end{equation}
which already implies soft-photon coherence and an AO pattern around charge-connected emitters~\cite{Ellis:1991qj}; hence, AO, or any variable monotonic with angle in the soft limit (e.g.\ $\pt$) with a global recoil scheme, reproduces LL (and NLL soft) structure without being strictly required by gauge dynamics. For EW emissions, gauge-boson masses screen the pure soft singularity; the relevant coherence is ordering in formation time,
$t_f\simeq 2\omega/(\pt^2+M_V^2)$, so an AO defined via a mass-aware angle $\theta_{\rm eff}^2\sim (\pt^2+M_V^2)/E^2$ (or, equivalently, a $\pt$-ordered scheme with an angular veto) captures the leading Sudakov structure at high scales where $\alpha_W\ln^2(Q/M_W)$ is sizable, while respecting SU(2)$\times$U(1) interference (isospin/charge flow) in both ISR and FSR~\cite{Dokshitzer:1991wu, Masouminia:2021kne}. Practically, we therefore (i) keep AO as the master clock to protect global QCD coherence; (ii) interleave QED and EW emissions on the same evolution scale with sector-appropriate kernels (charge/isospin dipoles), using a recoil strategy that preserves monotonicity with angle/formation time; and (iii) test numerically, using the envelope diagnostics of section~\ref{subsec:coh-summary}, whether adding QED/EW changes the wide-angle gluon-radiation pattern relative to the effect of changing the kinematic reconstruction scheme. In short, AO is enough for QED and EW showers provided it is \emph{mass-aware} for EW and embedded in an interleaved scheme that maintains the QCD angular hierarchy.

While preserving QCD coherence is primarily an algorithmic property (ensured by angular ordering and a monotonic soft-angle mapping), in practice predictions can be improved/restored by targeted tuning of shower and interleaving parameters against the observables of interest. In principle, such a programme may involve: (i) adopting the CMW rescaling of the coupling~\cite{Catani:1990rr} to incorporate coherent soft gluons beyond strict LL,
and tuning $\alphas(m_Z)$ and the shower cutoff $Q_0$; (ii) choosing and tuning the AO-compatible kinematic reconstruction scheme to stabilise the mapping $\tilde q\!\leftrightarrow\!\theta$ away from the strict soft-collinear limit; (iii) enforcing and, if needed, tuning angular vetoes/smooth-suppression parameters that protect the wide-angle region against overpopulation; (iv) controlling heavy-flavour mass effects (dead-cone realisation) via the heavy-quark mass scheme, $g\!\to\!Q\bar Q$ threshold handling, and quasi-collinear mass terms in the kernels; (v) interleaving QED/EW emissions on the same clock but tuning only sector-specific knobs (e.g.\ $\alphaem$ scale choice, EW Sudakov switches) while verifying that gluon wide-angle patterns remain unchanged; and (vi) constraining non-perturbative contaminations (MPI $p_{\text{T}0}$, colour reconnection, hadronization parameters) so they do not mimic or mask coherence. A tune should therefore be constrained by the observables studied here: LEP event and jet shapes, Tevatron and LHC radial energy profiles, and LHC jet masses, jet widths and leading-jet angularities, with high-scale or $100~\mathrm{TeV}$ projections used as stress tests of the same configurations.

In this study, we investigate the effects of interleaving QCD, QED, and EW radiation within an AO parton shower in Herwig~7. Herwig's AO parton shower supports three kinematic reconstruction schemes: $\pt$-preserving, virtuality-preserving and dot-product-preserving. To assess the effects of these interleaved parton showers, we switch between these schemes while activating different interactions within our shower ($\qcdonly$, $\qcdqed$, and $\qcdqedew$), focusing on jet shapes, event shapes and groomed-jet analyses based on LEP, Tevatron and LHC data. This provides a foundational basis for our results, with outcomes from Herwig's dipole shower included as a control comparison. Our primary objective is to understand how the interplay between electromagnetic and weak interactions with QCD affects the dynamics of successive radiations, potentially manifesting as changes in the overall shapes of events and/or jets.

This paper is organised as follows. In section~\ref{sec:aops} we review the AO shower in Herwig~7, detailing the evolution variable $\tilde q^2$ and its kinematic reconstruction schemes, recoil strategy, interleaving across QCD/QED/EW, and expected logarithmic accuracy/coherence properties. In section~\ref{sec:simu}, we present the simulation setup and results for the observables in this study: LEP, Tevatron, and LHC event shapes; groomed and ungroomed jet masses; jet widths; radial jet shapes; and leading-jet angularities, along with the corresponding $100~\mathrm{TeV}$ projections. We compare combinations of AO kinematic reconstruction schemes and shower-sector choices. In section~\ref{sec:tune}, we describe our tuning procedure and constraints, and validate the post-tune predictions. Finally, in section~\ref{sec:conc} we summarise our findings and outline future directions.


\section{Shower Kinematics and Kernels}
\label{sec:aops}

In the soft and collinear limits of phase space, the probability of an emission of a gluon from a quark $\diffd P$ can be written as 
\begin{equation}
    \diffd P = \frac{\CF \alphas(\ptsq)}{\pi} \frac{\diffd \omega}{\omega} \frac{\sin\theta \, \diffd \theta}{1-\cos\theta} \Theta(\theta_{\text{prev}} - \theta)\, ,
\end{equation} 
where $\CF$ is the colour factor, $\alphas$ is the strong coupling constant, $\omega$ is the energy of the emitted gluon, and $\theta$ is the emission angle relative to the parent parton's direction. The Heaviside step function $\Theta(\theta_{\text{prev}} - \theta)$ enforces angular ordering by restricting subsequent emissions to smaller angles, $\theta<\theta_{\text{prev}}$, where $\theta_{\text{prev}}$ denotes the angle of the preceding emission. 
It is useful to trade the variables $(\omega,\theta)$ for shower variables $(z,\tilde q^2)$. In the Herwig AO shower, $z$ is defined as a light-cone momentum fraction rather than simply an energy fraction. Hence, the relation between relative transverse momentum, emission angle and energy sharing is unambiguous only in the strict massless soft-collinear limit. Away from that limit, it depends on the treatment of masses, daughter virtualities, and recoil, motivating the kinematic reconstruction schemes discussed below. For an on-shell massless final-state branching, the AO variable reduces to
\begin{equation}
    \tilde{q}^2 = \frac{\ptsq}{z^2 (1-z)^2}\, ,
    \label{EV}
\end{equation}
with $z$ being the light-cone momentum fraction that parametrises the emitting quark's post-emission momentum. In the small-angle limit this corresponds to $\tilde q\simeq E\,\theta$, so that assigning daughter starting scales $\tilde q_1<z\tilde q$ and $\tilde q_2<(1-z)\tilde q$ implements angular ordering for subsequent branchings.

For massive partons, eq.~\eqref{EV} can be written as
\begin{equation}
\label{eq:massive-qtilde}
\tilde{q}^2 =
\frac{
q^2-m^2
}{
z(1-z)
}
=
\frac{
2q_1\!\cdot\!q_2+m_1^2+m_2^2-m^2
}{
z(1-z)
}
=
\frac{
\ptsq-z(1-z)m^2+(1-z)m_1^2+z m_2^2
}{
z^2(1-z)^2
}\, ,
\end{equation}
where $m$ is the on-shell mass of the parent, and $m_1$ and $m_2$ are the corresponding on-shell masses of the two children. When the child partons are on-shell, all three forms on the right-hand side of eq.~\eqref{eq:massive-qtilde} are equal. With this evolution variable, angular ordering is defined as $\tilde{q}_{i+1}<z\tilde{q}_i$.

The emission probability can be rewritten as
\begin{equation}
\diffd P =
\frac{\alphas\!\left(z^2(1-z)^2\tilde{q}^2\right)}{2\pi}
\frac{\diffd\tilde{q}^2}{\tilde{q}^2}
\diffd z\, P_{qq}\,
\Theta(\tilde{q}_{i+1}<z\tilde{q}_i)\, ,
\end{equation}
where $P_{qq}$ is the quasi-collinear splitting function, valid in the limit where $\ptsq$ and the masses are small compared with the hard scale, while ratios such as $\ptsq/m^2$ are kept finite~\cite{Gieseke:2003rz}.

After subsequent showering, the children are no longer necessarily on shell. We denote their four-momenta by $q_1^\mu$ and $q_2^\mu$, so that $q_i^2$ is the daughter virtuality after its own emissions, while $m_i$ remains the corresponding on-shell mass parameter entering the splitting kernel. The three forms on the right-hand side of eq.~\eqref{eq:massive-qtilde} are then no longer equivalent. We can choose any one of the quantities $\ptsq$, $q^2$ or $q_1\!\cdot\!q_2$, calculate it from eq.~\eqref{eq:massive-qtilde} and then, after showering the children, reconstruct their momenta according to that value.

In the $\pt$-preserving reconstruction scheme, $\ptsq$ is calculated as
\begin{equation}
\ptsq =
z^2(1-z)^2\tilde{q}^2
+z(1-z)m^2
-(1-z)m_1^2
-zm_2^2\, .
\end{equation}

In the virtuality-preserving reconstruction scheme, the transverse momentum is calculated from the generated virtuality, replacing the on-shell masses in the single-emission expression by the daughter virtualities,
\begin{equation}
\ptsq =
z^2(1-z)^2\tilde{q}^2
+z(1-z)m^2
-(1-z)q_1^2
-zq_2^2\, .
\end{equation}
This choice can generate unphysical configurations with $\ptsq<0$ after subsequent emissions; in that case, the transverse momentum is frozen to zero~\cite{Reichelt:2017hts}.

The dot-product-preserving reconstruction scheme instead keeps the relation between the AO evolution variable and the dot product of the two daughter momenta~\cite{Marchesini:1987cf, Bewick:2019rbu}. With the above notation,
\begin{equation}
\ptsq =
z^2(1-z)^2\tilde{q}^2
-(1-z)^2q_1^2
-z^2q_2^2
+z(1-z)\bigl(m^2-m_1^2-m_2^2\bigr)\, .
\end{equation}
For gluon emission, a sufficient, but not necessary, condition for the dot-product-preserving reconstruction to give a physical transverse momentum is~\cite{Bewick:2019rbu}
\begin{equation}
\tilde{q}^2>
2\max\left(
\frac{q_1^2}{z^2},
\frac{q_2^2}{(1-z)^2}
\right)\, .
\end{equation}
When this condition fails, $\ptsq$ can still become negative, for example for massive children such as heavy quarks or electroweak bosons, but much less often than in the virtuality-preserving scheme; such cases are treated in the same way.
For $g\to Q\bar Q$ branchings, the massive threshold condition is symmetric under $z\leftrightarrow 1-z$. Setting $m=0$ and $m_1=m_2=m_Q$ in the on-shell expression gives
\begin{equation}
\ptsq=z^2(1-z)^2\tilde q^2-m_Q^2\, .
\end{equation}
Thus the physical region requires
\begin{equation}
z^2(1-z)^2\tilde q^2 \geq m_Q^2\, ,
\end{equation}
or, if the shower cutoff is imposed explicitly,
\begin{equation}
z^2(1-z)^2\tilde q^2 \geq m_Q^2+\ptsqmin\, .
\end{equation}
Equivalently, an allowed value of $z$ exists only for $\tilde q^2\geq 16m_Q^2$, or $\tilde q^2\geq 16(m_Q^2+\ptsqmin)$ when the cutoff is included. A thorough analysis of the logarithmic accuracy of these choices was made in~\cite{Bewick:2019rbu}.
The dot-product-preserving scheme was found to overestimate the rate of $e^+e^-\to q\bar{q}$ events in which the jets acquire significant virtuality due to multiple emissions of similar hardness, rather than a single hard emission. In~\cite{Bewick:2019rbu}, a correction for this was introduced, in which events are accepted with probability proportional to the two-body phase-space volume for decay to particles of the generated invariant masses. Otherwise, the generated showers are vetoed and new showers generated. For this reason, this option is often called the dot-product$+$veto scheme in Herwig.

All possible splittings in the shower can be categorised by their interaction type. The QCD emissions include
\begin{equation}
    q \to qg, \quad g \to gg, \quad g \to q\bar{q}\, ,
\end{equation}
and are generated using quasi-collinear splitting functions,
\begin{equation}
\begin{aligned}
P_{qq} (z, \tilde{q}) &= \CF \frac{1}{1-z} \left( 1 + z^2 - \frac{2m^2}{z \tilde{q}^2}\right), \\
P_{gg} (z, \tilde{q}) &= \CA \left( \frac{z}{1-z} + \frac{1-z}{z} + z(1-z)\right), \\
P_{gq} (z, \tilde{q}) &=\TR \left( 1 -2z(1-z) + \frac{2m^2}{z (1-z) \tilde{q}^2}\right)\, ,
\label{LO-spltFn}
\end{aligned}
\end{equation}
with $\CF = 4/3$, $\CA = 3$ and $\TR = 1/2$. The QED emissions allow quarks to radiate photons, which can then split into charged fermion pairs,
\begin{equation}
    q \to q \gamma, \quad \gamma \to f\bar{f}\, ,
\end{equation}
and their splitting functions are defined by the corresponding QCD kernels, with $\alphas$ replaced by $\alphaem$ and the colour factors by the appropriate electromagnetic factors. For photon emission from a quark, this corresponds to $\CF\to Q_q^2$, while for photon splitting one has $\TR\to \encee^f Q_f^2$, where $Q_f$ is the electric charge of the produced fermion and $\encee^f=3$ for quarks and $\encee^f=1$ for leptons.
The EW splittings derived and defined in~\cite{Masouminia:2021kne} incorporate the radiation of $W$, $Z$ and $H$ bosons from quarks
\begin{equation}
q \to q'W^{\pm}, \quad q \to qZ^0, \quad q \to qH\, ,
\end{equation}
as well as final-state emissions from the bosons themselves,
\begin{equation}
\begin{gathered}
W^{\pm} \to W^{\pm}Z^0, \quad W^{\pm} \to W^{\pm}\gamma, \quad W^{\pm} \to W^{\pm}H, \\
Z^0 \to W^+W^-, \quad \gamma \to W^+W^-,  \quad Z^0 \to Z^0H\, .
\end{gathered}
\end{equation}
The splitting functions in this case are defined using helicity amplitudes,
\begin{equation}
\begin{aligned}
P_{\widetilde{ij}\to i+j}(z,\tilde{q})
&= \sum_{\lambda_0,\lambda_1,\lambda_2}
\left|H_{\widetilde{ij}\to i+j}(z,\tilde{q};\lambda_0,\lambda_1,\lambda_2)\right|^2 \\
&= \sum_{\lambda_0,\lambda_1,\lambda_2}
\left|g\,F_{\lambda_0,\lambda_1,\lambda_2}^{\widetilde{ij}\to i+j}\right|^2\, .
\end{aligned}
\end{equation}
Here $\lambda_0$ denotes the helicity of the parent particle, while $\lambda_1$ and $\lambda_2$ denote the helicities of the two children. The displayed expression is the helicity-summed kernel; when the shower retains helicity information, the corresponding fixed-$\lambda_0$ kernel is used before summing or averaging over the parent spin state. For an unpolarised parent, the usual average over the allowed parent helicities is understood in the spin-averaged shower rate. The factor $g$ denotes the relevant gauge or Yukawa coupling of the vertex. Process-dependent charge, weak-isospin, chirality, CKM and colour factors are included in the helicity vertex function $F^{\widetilde{ij}\to i+j}_{\lambda_0,\lambda_1,\lambda_2}$; for QCD branchings these reduce, after the standard colour sums, to the usual colour factors appearing in eq.~\eqref{LO-spltFn}, while for QED and EW branchings they give the corresponding charge and weak-isospin factors. The use of this helicity information within the spin-density-matrix formalism, including its application to polarised deep-inelastic scattering in Herwig~7, is discussed in~\cite{Masouminia:2026bhk}. A gauge-invariant treatment of the longitudinal electroweak modes has more recently been formulated in~\cite{Masouminia:2026udz}. That construction provides an alternative longitudinal basis for the EW shower, based on the Dawson remainder supplemented by a Ward-identity-fixed Goldstone matching term. It is not used in the present study, which retains the default EW shower setup described above. This does not affect the interleaving comparisons presented here.


\section{Testing the Effect of Interleaving}
\label{sec:simu}

In the previous section, we summarised the three kinematic reconstruction schemes and the splitting functions for QCD, QED, and EW splittings. QCD splittings are always included, and whenever EW splittings are enabled, we also include QED splittings, giving nine possible shower settings. For user-facing runs, the default from Herwig~7.4 onward will be the dot-product scheme with all splittings active. In this study, however, we evaluate all combinations within the same simulation and analysis and assess the resulting differences.

To validate QCD structure in our AO parton shower, we study a selection of IRC-safe observables at LEP, Tevatron and the LHC, together with corresponding projections to $\sqrt{s}=100~\mathrm{TeV}$: event shapes ($1\!-\!T$, $C$, $B_T$, $y_{23}^{\text{Durham}}$), groomed and ungroomed jet masses, jet widths, leading-jet angularities, and radial jet-shape profiles ($\rho(r),\,\Psi(r)$)~\cite{Ellis:1991qj, Dokshitzer:1991wu, Dasgupta:2003iq}. For each setup, we keep PDFs, scales, hadronization, and tune identical, and compare to (a) LEP data at $M_Z$ and 130--209~GeV, (b) Tevatron and LHC fiducial measurements, and (c) $100~\mathrm{TeV}$ projections using the same observable definitions. The $100~\mathrm{TeV}$ panels are prediction-only reruns of the same generated process and the same \textsc{Rivet} analysis used for the corresponding data comparison, with only the collider energy changed; no comparison to data is implied. In these panels, the ratios are taken relative to the dipole shower prediction at the same energy, as an independent baseline. Both the original and the $100~\mathrm{TeV}$ simulations were analysed using the following \textsc{Rivet} analyses:

\begin{enumerate}

\item ALEPH (\href{https://rivet.hepforge.org/analyses/ALEPH_2004_I636645.html}{\texttt{ALEPH\_2004\_I636645}}) $e^+e^-$ event shapes and jet rates across multiple $\sqrt{s}$ values~\cite{ALEPH:2003obs}. We use global event shapes ($1{-}T$, $C$, $B_{T, W}$, $M_H^2/s$), Durham $y_{23}$ and $y_{34}$, and charged particle momentum distributions, with standard ALEPH hadronic selections at the $Z$ pole and LEP\,2 (130--209\,GeV), as well as $S$, $A$, $O$ and $M_L^2/s$ and higher jet rates at the $Z$ pole. These are classic probes of QCD shower dynamics.

\item CDF (\href{https://rivet.hepforge.org/analyses/CDF_2005_I682179.html}{\texttt{CDF\_2005\_I682179}}) jet shapes at $\sqrt{s}=1.96$\,TeV~\cite{CDF:2005prv}. Differential and integrated jet shapes $\rho(r),\Psi(r)$ are taken for central jets in Tevatron $p\bar p$ data using the cone jet algorithm with $R=0.7$. Large-$r$ energy flow constrains wide-angle soft radiation, recoil, and heavy-flavour dead-cone modelling.

\item ATLAS (\href{https://rivet.hepforge.org/analyses/ATLAS_2011_I882984.html}{\texttt{ATLAS\_2011\_I882984}}) jet shapes at 7\,TeV~\cite{ATLAS:2011bvd}. Differential and integrated jet shapes for inclusive anti-$\kt$ jets with $R=0.6$.

\item ATLAS (\href{https://rivet.hepforge.org/analyses/ATLAS_2012_I1119557.html}{\texttt{ATLAS\_2012\_I1119557}}) jet substructure observables at 7\,TeV~\cite{ATLAS:2012nnf}. Jet mass, jet width, planar flow and angularity, for high-$\pt$
jets ($\pt>300$\,GeV) reconstructed using the anti-$\kt$ algorithm with $R=0.6$ and $R=1.0$.

\item ATLAS (\href{https://rivet.hepforge.org/analyses/ATLAS_2013_I1243871.html}{\texttt{ATLAS\_2013\_I1243871}}) jet shapes in $t\bar t$ at 7\,TeV~\cite{ATLAS:2013uet}. Differential and integrated jet shapes for $b$-tagged and light-quark jets in $t\bar t$ lepton+jets events, using the anti-$\kt$ algorithm with $R=0.4$. Radiation off heavy colour sources probes mass effects (dead cone) within AO and their interplay with recoil and evolution mapping.

\item CMS (\href{https://rivet.hepforge.org/analyses/CMS_2013_I1224539.html}{\texttt{CMS\_2013\_I1224539}}) jet masses in $W$, $Z$, and dijet events at 7\,TeV~\cite{CMS:2013kfv}. Groomed and ungroomed jet-mass spectra in $V{+}$jets and dijets with standard selections. Ungroomed jet-mass spectra are especially sensitive to soft wide-angle emissions and test whether AO interleaving leaves the expected angular structure unchanged, while groomed jet masses primarily probe hard emission within the jets.

\item CMS (\href{https://rivet.hepforge.org/analyses/CMS_2018_I1682495.html}{\texttt{CMS\_2018\_I1682495}}) ungroomed and SoftDrop-groomed jet-mass distributions in dijet events at 13\,TeV~\cite{CMS:2018vzn}. We use the ungroomed and groomed mass distributions to compare the sensitivity of inclusive and groomed jet structure to the kinematic reconstruction scheme and shower-sector choice.

\end{enumerate}
While we generated results for all these analyses, and will use them for tuning in the next section, we discuss observables of interest from just some of them in the subsections below. Details of the shower dynamics are tested by checking that wide-angle soft radiation is not overpopulated in the displayed event-shape, jet-mass, jet-width, angularity and radial-profile observables, and that adding QED or EW shower sectors does not produce a larger distortion of the gluon-radiation pattern than changing the kinematic reconstruction scheme itself. Discrepancies across kinematic reconstruction schemes quantify their impact beyond the strict soft-collinear limit, while discrepancies across interleaving choices diagnose possible double-counting or loss of destructive interference. Mass effects and heavy-flavour thresholds can also affect several of these observables by altering shower dynamics. This is summarised in section~\ref{subsec:coh-summary} by a quantitative comparison of interleaving-induced and scheme-induced envelopes, which then guides the dedicated tune and the choice of preferred kinematic reconstruction scheme and interleaving strategy~\cite{Bellm:2019zci, Bewick:2023tfi}.

\subsection{Thrust}

Thrust measures how pencil-like the final state is and, in the two-jet limit, is sensitive to soft radiation over a wide angular range. Since a small-angle splitting should not create an additional independent source of radiation at larger angles, thrust is a useful probe of the extent to which the interleaved splittings preserve this. Thrust is defined as
\begin{equation}
T \;=\; \max_{\hat{\mathbf n}\in S^2}
\frac{\displaystyle \sum_i \big|\mathbf p_i \!\cdot\! \hat{\mathbf n}\big|}
     {\displaystyle \sum_i \big|\mathbf p_i\big|}\,,
\qquad \text{with the thrust axis } \hat{\mathbf n}_T \text{ attaining the maximum}\, .
\end{equation}
One typically studies $1{-}T\ll 1$ in the dijet regime. For a single soft emission of energy $\omega$ at polar angle $\theta$ relative to $\hat{\mathbf{n}}_T$, the leading-power contribution is
\begin{equation}
  1{-}T \;\simeq\; \frac{\omega}{Q}\left(1-|\cos\theta|\right)
  \;=\; \frac{k_\perp}{Q}e^{-|\eta|}\, ,
\end{equation}
where $k_\perp=\omega\sin\theta$ and $|\eta|=-\ln\tan(\theta/2)$ is measured from the thrust axis in the relevant hemisphere. In the collinear limit this gives $1{-}T\simeq \omega\theta^2/(2Q)$. More generally, in the two-jet limit,
\begin{equation}
  1{-}T \;\simeq\; \frac{1}{Q}\sum_{\text{emissions}} k_\perp\,e^{-|\eta|}\, ,
\end{equation}
so $1{-}T$ is a global, IRC-safe measure of angle-weighted transverse-momentum flow~\cite{Dasgupta:2003iq}.
Coherence implies that radiation at angles larger than a collinear parton system resolves only its total colour charge; a subsequent small-angle splitting therefore does not change the large-angle radiation rate. In AO showers, this structure is implemented by generating wide-angle emissions before subsequent, more collinear branchings. The $1{-}T\to0$ region consequently exhibits the standard Sudakov suppression associated with the absence of emissions giving sizeable $k_\perp e^{-|\eta|}$. Enhancement of wide-angle radiation would shift the peak position; we therefore check that QED and EW interleaving leaves the wide-angle gluon pattern unchanged.

\begin{figure}[t]
\centering
\includegraphics[width=.49\textwidth]{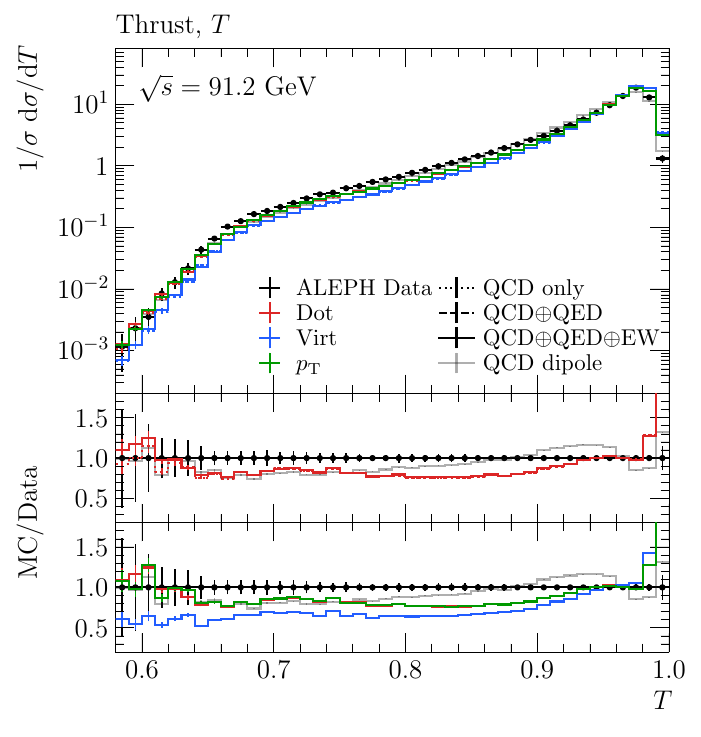}
\includegraphics[width=.49\textwidth]{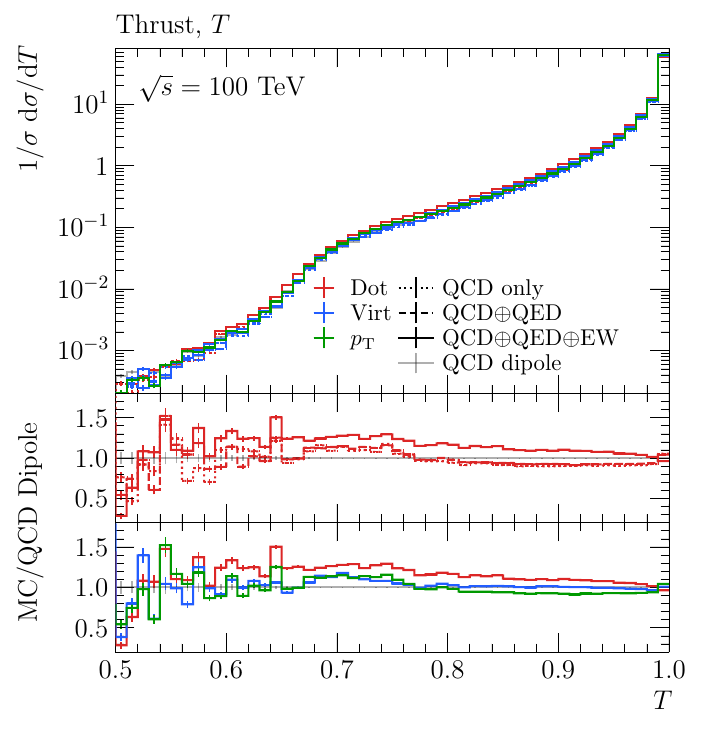}
\caption{Thrust in $e^+e^-$ collisions. The left panel shows ALEPH measurements at $E_{\rm CMS}=91.2~\text{GeV}$~\cite{ALEPH:2003obs}; the right panel gives predictions at $E_{\rm CMS}=100~\text{TeV}$ using the same observable definition. Ratio panels compare MC to data (left) and to the QCD dipole shower, which serves as a common MC reference (right). The upper ratio plots show all three interleaving choices with the dot-product-preserving reconstruction scheme, and the lower ones show all three schemes with the $\qcdqedew$ interleaved shower. In the two-jet limit ($1{-}T\!\ll\!1$), the coherent angular structure yields a steep Sudakov fall-off, while an enhancement of wide-angle radiation would harden the tail. }
\label{fig:ALEPH_2004_Thrust}
\end{figure}

Figure~\ref{fig:ALEPH_2004_Thrust} uses thrust {in $e^+e^-$ annihilation} as a benchmark of QCD shower dynamics. The left panel shows ALEPH measurements at $E_{\rm CMS}=91.2~\text{GeV}$, while the right panel shows the corresponding predictions at $E_{\rm CMS}=100~\text{TeV}$. Curves compare the AO shower for three kinematic reconstruction schemes (dot-product$+$veto ``Dot'', virtuality ``Virt'', and $\pt$ preserving) and three interleaving choices ($\qcdonly$, $\qcdqed$, $\qcdqedew$), together with a dipole-shower baseline. The distributions and ratios therefore directly test whether the QCD structure is preserved.

\subsection[$C$-parameter]{\boldmath $C$-parameter}

The $C$-parameter quantifies the dispersion of momentum flow over solid angle. Since a small-angle splitting should not generate an additional independent source at larger angles, an enhancement of wide-angle radiation would populate the shoulder region. The $C$-parameter is defined from the linearised momentum tensor
\begin{equation}
  \Theta^{\alpha\beta} \;=\; 
  \frac{\displaystyle \sum_i \frac{p_i^\alpha p_i^\beta}{|\mathbf{p}_i|}}
       {\displaystyle \sum_i |\mathbf{p}_i|}\, ,
\end{equation}
with eigenvalues $\{\lambda_1,\lambda_2,\lambda_3\}$, and
\begin{equation}
  C \;=\; 3\!\left(\lambda_1\lambda_2+\lambda_2\lambda_3+\lambda_3\lambda_1\right),
  \qquad 0 \le C \le 1\, .
\end{equation}
For a single soft emission of energy $\omega$ at polar angle $\theta$ relative to the thrust axis, one finds at leading power
\begin{equation}
  C \;\simeq\; 3\,\frac{\omega}{Q}\,\sin^2\theta
  \;\approx\; 3\,\frac{\omega}{Q}\,\theta^{2}\, ,
\end{equation}
in the collinear limit. Compared with thrust, for which $1{-}T\simeq \omega\theta^2/(2Q)$ in the same limit, the $C$-parameter has the same soft-collinear angular scaling. Radiation at angles larger than a collinear parton system resolves only its total colour charge, so subsequent small-angle branchings should not increase the large-angle radiation rate. Angular ordering implements this separation of angular regions. The small-$C$ region consequently exhibits the standard Sudakov suppression associated with the absence of emissions giving sizeable energy-weighted $\sin^2\theta$. Overpopulating the wide-angle region would harden the spectrum and shift probability into the shoulder. We test this by comparing the small-$C$ resummation region and the hard-emission shoulder across kinematic reconstruction schemes and interleaving choices.

\begin{figure}[t]
\centering
\includegraphics[width=.49\textwidth]{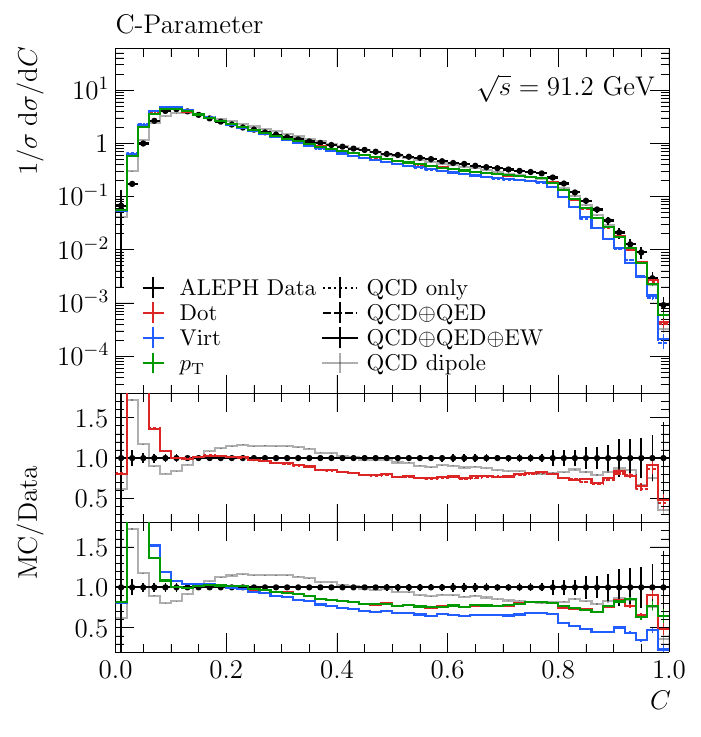}
\includegraphics[width=.49\textwidth]{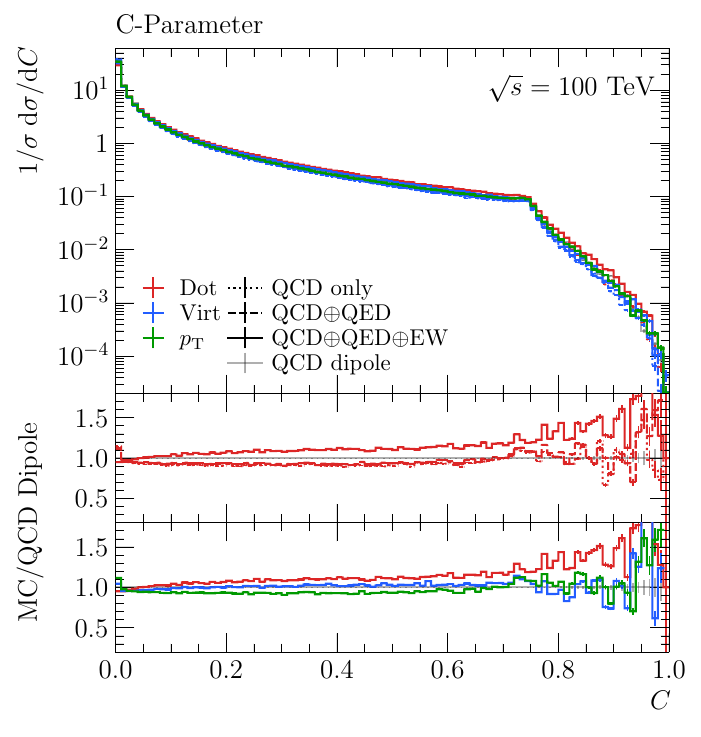}
\caption{$C$-parameter in $e^+e^-$ collisions. The notation follows that of figure~\ref{fig:ALEPH_2004_Thrust}.}
\label{fig:ALEPH_2004_C}
\end{figure}

Figure~\ref{fig:ALEPH_2004_C} focuses on the $C$-parameter. Using the same setup and notation as figure~\ref{fig:ALEPH_2004_Thrust}, we compare predictions to ALEPH data at $E_{\rm CMS}=91.2~\text{GeV}$ and present predictions at $100~\text{TeV}$. The $C$ spectrum complements our probe of soft radiation at sizeable angles: the small-$C$ region is controlled by Sudakov suppression of emissions with sizeable energy-weighted $\sin^2\theta$. In contrast, an enhancement of large-angle radiation would populate the region towards the three-jet shoulder.

\subsection[Total Jet Broadening $B_T$]{\boldmath Total jet broadening $B_T$}

Total jet broadening $B_T$ measures the transverse spread about the thrust axis and is linearly sensitive to $k_\perp$, providing a complementary probe of recoil and soft wide-angle radiation.
Define hemispheres $H_{1,2}$ with respect to the thrust axis $\hat{\mathbf{n}}_T$ and
\begin{equation}
  B_k \;=\; \frac{1}{2\sum_i |\mathbf{p}_i|}\sum_{i\in H_k} \big|\mathbf{p}_i \times \hat{\mathbf{n}}_T\big| \,,
  \qquad k=1,2, \qquad B_T \equiv B_1 + B_2\, .
\end{equation}
For a single soft emission of energy $\omega$ at polar angle $\theta$, one has $B_T\sim k_\perp/(2Q)$ with $k_\perp\simeq\omega\theta$, so small $B_T$ is Sudakov-suppressed. Its linear sensitivity to transverse momentum makes it responsive to changes in recoil and soft radiation, although it is not the most sensitive of the event-shape observables considered here.
\begin{figure}[t]
\centering
\includegraphics[width=.49\textwidth]{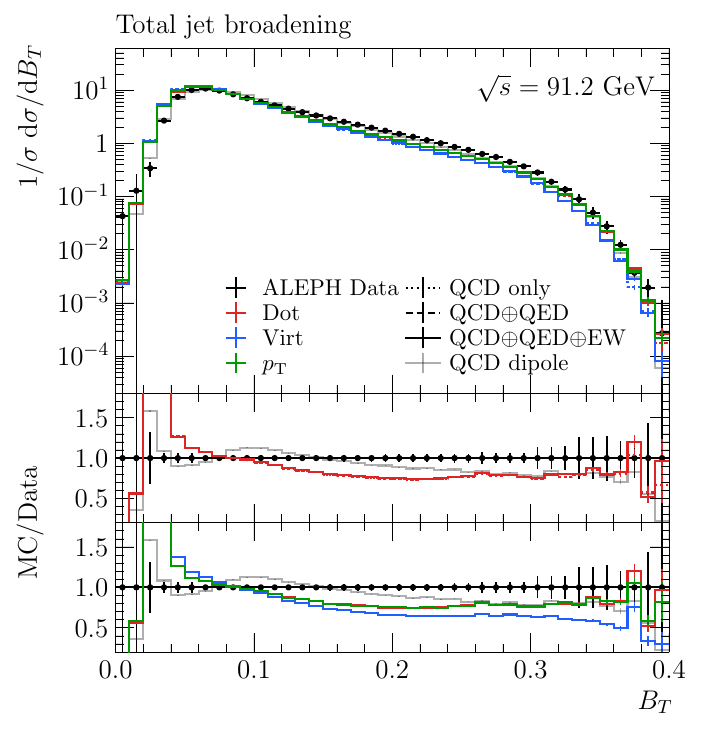}
\includegraphics[width=.49\textwidth]{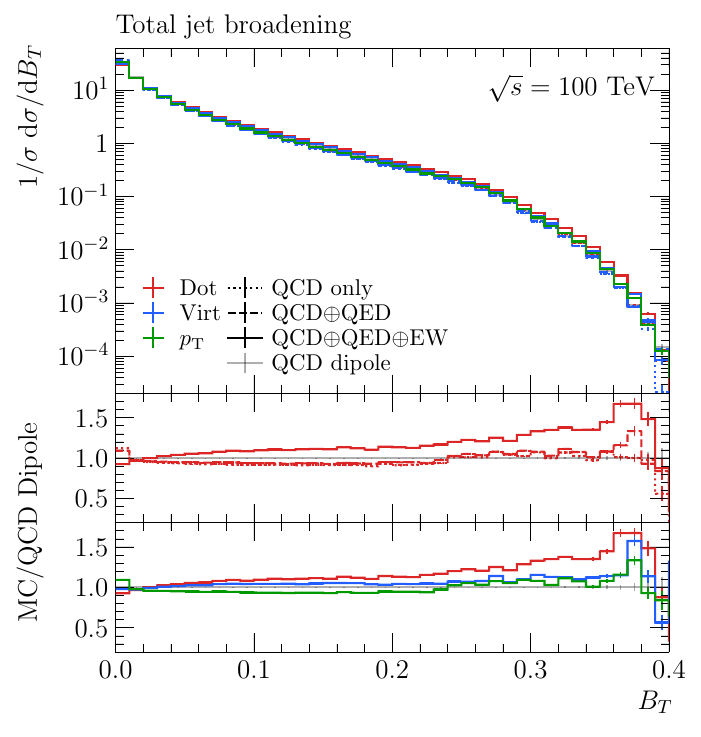}
\caption{Total jet broadening $B_T$ in $e^+e^-$ collisions. The notation follows that of figure~\ref{fig:ALEPH_2004_Thrust}.}
\label{fig:ALEPH_2004_Broadening}
\end{figure}

Figure~\ref{fig:ALEPH_2004_Broadening} highlights the additional leverage of $B_T$ on recoil and wide-angle soft radiation. Stability of the small-$B_T$ Sudakov region across all showers indicates that the angular radiation pattern is retained, whereas recoil artefacts or competition between shower sectors would enhance the mid-$B_T$ yield.

\subsection[Durham Transition Scale $y_{23}$]{\boldmath Durham transition scale $y_{23}$}

The Durham resolution directly tracks the scale of the first resolvable emission. Clustering with the Durham metric
\begin{equation}
  y_{ij} \;=\; \frac{2\,\min(E_i^2,E_j^2)\,\bigl(1-\cos\theta_{ij}\bigr)}{s}
  \;\simeq\; \frac{k_{\perp,ij}^{2}}{Q^{2}} \quad (\text{soft-collinear})\, ,
\end{equation}
one iteratively merges the pair with minimum $y_{ij}$ until two jets remain; the \emph{transition scale} $y_{23}$ is the value of $y_{\rm cut}$ at the $3\!\to\!2$ merge. In the two-jet limit, $y_{23}$ is controlled by the primary Sudakov form factor and is therefore similar in spirit to event-shape observables such as $B_T$, while being tied directly to the first resolved Durham emission scale. Overcounting wide-angle soft radiation would increase the three-jet rate at fixed $y_{\rm cut}$, harden the distribution, and distort its Sudakov slope.
\begin{figure}[t]
\centering
\includegraphics[width=.49\textwidth]{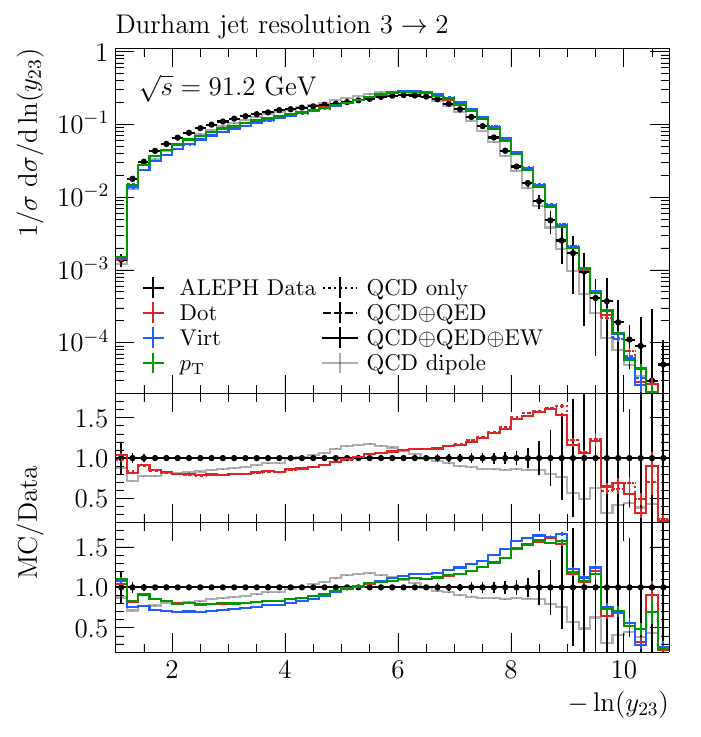}
\includegraphics[width=.49\textwidth]{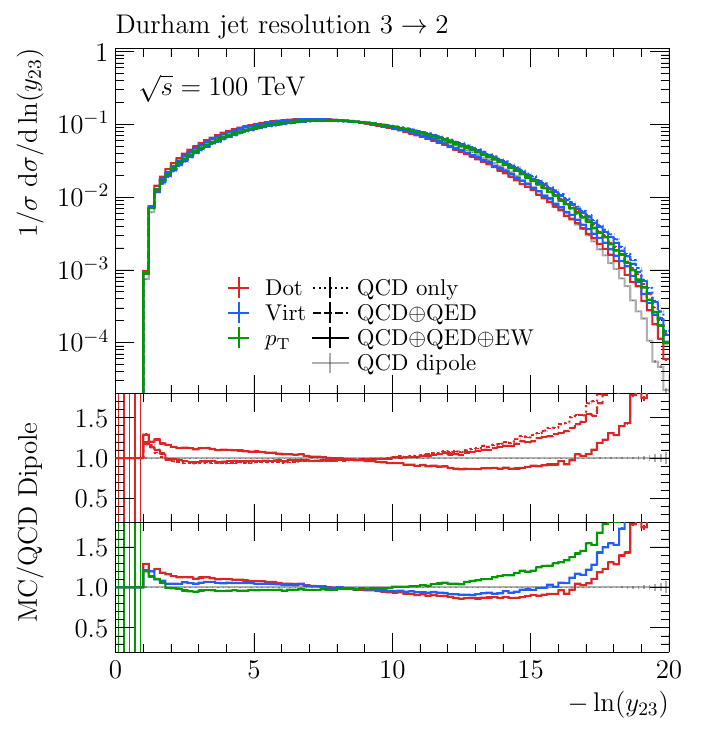}
\caption{Durham $3\!\to\!2$ transition scale $y_{23}$ in $e^+e^-$ collisions. The notation follows that of figure~\ref{fig:ALEPH_2004_Thrust}.}
\label{fig:ALEPH_2004_y23}
\end{figure}
Figure~\ref{fig:ALEPH_2004_y23} highlights $y_{23}$ as a diagnostic using logarithmic binning in the Durham transition scale, shown as $-\ln y_{23}$. This resolves the two-jet Sudakov region over several orders of magnitude and makes changes in the Sudakov slope more visible.

Taken together, figures~\ref{fig:ALEPH_2004_Thrust}, \ref{fig:ALEPH_2004_C}, \ref{fig:ALEPH_2004_Broadening} and~\ref{fig:ALEPH_2004_y23} show a consistent pattern across the $e^+e^-$ event-shape observables. Compared with the LEP data, the virtuality-preserving scheme yields less radiation in the hard-emission regions than the dot-product- and $\pt$-preserving schemes, which remain close to one another. Activating QED and EW emissions produces no significant change. The same suppression in the virtuality-preserving scheme persists in the 100~TeV projections. QED radiation again has little visible effect, while including EW emissions gives a small additional contribution in parts of the distributions.

\subsection{Jet Mass}

Jet mass probes the aggregate soft-collinear activity captured by a jet and is therefore a direct diagnostic of how a shower populates wide angles inside a fixed radius. For a jet reconstructed from constituents $i$ with four-momenta $p_i^\mu$, the invariant mass is
\begin{equation}
m_J^2 \;=\; \Big(\sum_i E_i\Big)^2 \;-\; \Big|\sum_i \mathbf{p}_i\Big|^2\, ,
\end{equation}
and, for a single emission with momentum fraction $z$ at angle $\theta$ to the jet axis,
\begin{equation}
m_J^2 \;\simeq\; z(1-z)\,p_{\text{T},J}^2\,\theta^2 \;\xrightarrow{\,z\ll1\,}\; z\,p_{\text{T},J}^2\,\theta^2, 
\qquad 
\rho \;\equiv\; \frac{m_J^2}{p_{\text{T},J}^2 R^2}\, ,
\end{equation}
where $\pt$ is the jet's transverse momentum, so the $m_J$ (or $\rho$) spectrum is set by the interplay of soft transverse momentum and emission angle up to the jet radius $R$. Since radiation at angles larger than a collinear parton system resolves only its total colour charge, subsequent small-angle branchings should not create additional independent radiation near the jet boundary. Overpopulating this region through reconstruction or interleaving would harden the high-mass tail. The ATLAS measurement therefore provides a clean, infrared-safe test of these mass-systematics effects~\cite{ATLAS:2012nnf}.

\begin{figure}[t]
\centering
\includegraphics[width=.49\textwidth]{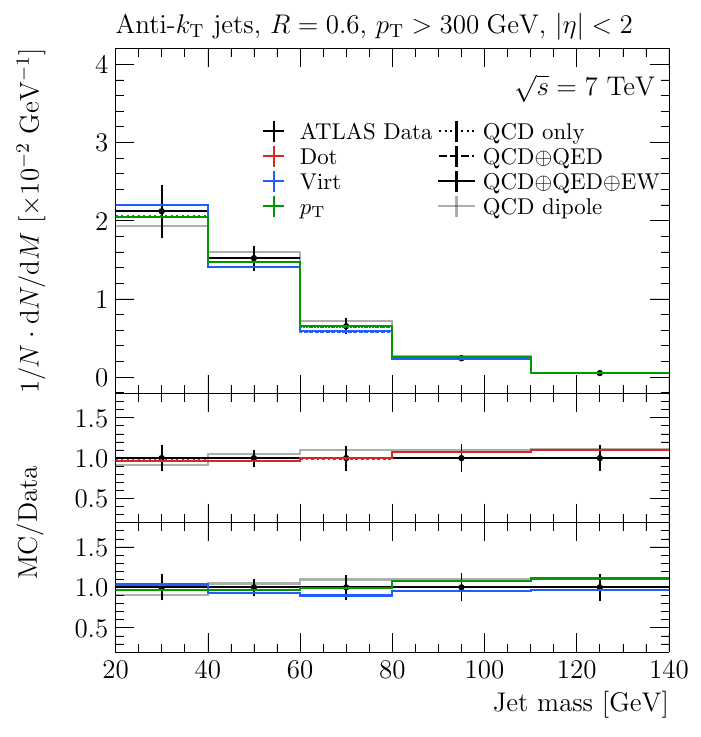}
\includegraphics[width=.49\textwidth]{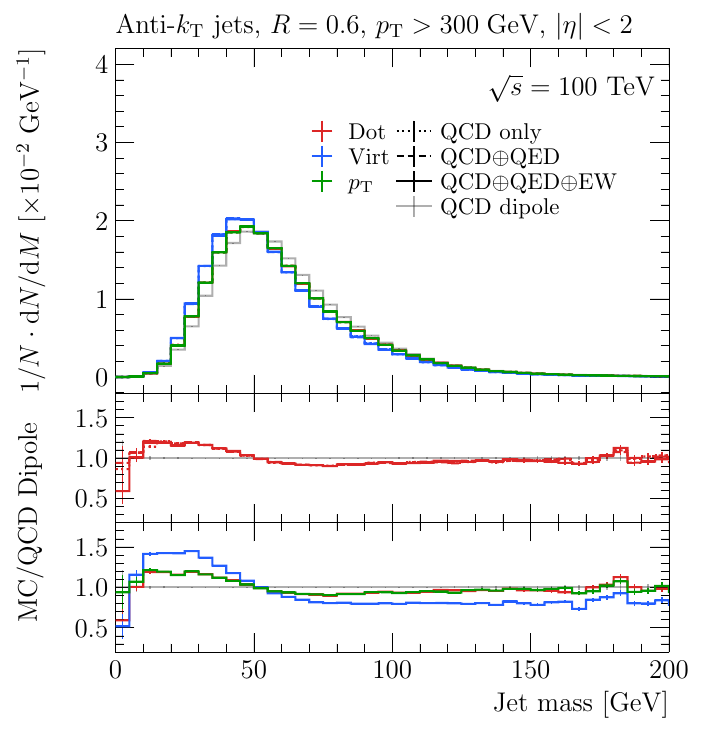}
\caption{Jet mass for leading anti-$\kt$ jets. Left: ATLAS measurement at $\sqrt{s}=7~\mathrm{TeV}$ with $\pt>300~\mathrm{GeV}$, $|\eta|<2$, and $R=0.6$, corrected to particle level~\cite{ATLAS:2012nnf}. Right: particle-level predictions at $E_{\rm CMS}=100~\mathrm{TeV}$. The small-$m_J$ Sudakov region and high-mass tail test whether the reconstruction or interleaving procedure overpopulates wide-angle radiation near the jet boundary. The notation follows that of figure~\ref{fig:ALEPH_2004_Thrust}.}
\label{fig:ATLAS_2012_JetMass}
\end{figure}

Figure~\ref{fig:ATLAS_2012_JetMass} shows the leading-jet mass distribution for the ATLAS selection and compares how it changes under the different kinematic reconstruction schemes and interleaving choices. The low-mass region tests the Sudakov suppression of emissions contributing to $m_J$, while the high-mass tail is sensitive to harder or wider-angle radiation inside the jet radius.
Ungroomed jet mass is sensitive to soft emissions over the full jet radius, so its small-$\rho$ Sudakov region and tail test whether the reconstruction or interleaving procedure increases wide-angle radiation. Grooming removes soft, large-angle contributions and isolates the hardest quasi-collinear branch, sharpening the connection between the measured mass and the primary emission in the shower. We adopt SoftDrop (mMDT for $\beta\!=\!0$~\cite{Dasgupta:2013via,Larkoski:2014wba}) on the Cambridge-Aachen declustering tree with the condition
\begin{equation}
  \frac{\min(p_{\text{T}1},p_{\text{T}2})}{p_{\text{T}1}+p_{\text{T}2}} \;\equiv\; z \;>\; z_{\rm cut}\,\Big(\tfrac{\theta}{R}\Big)^{\beta},
  \qquad \beta\ge 0\, ,
\end{equation}
so that, for $\beta\!=\!0$, purely soft wide-angle emissions are eliminated and the groomed mass is set by the first emission passing the cut, $m_{J,{\rm groom}}^2 \simeq z\,\ptsq\,\theta^2$ with $z>z_{\rm cut}$. The groomed mass has a calculable Sudakov structure and reduced sensitivity to non-perturbative effects, while overpopulation of large-angle radiation would be more directly visible in the ungroomed mass through a hardened tail or shifted peak. The joint comparison of groomed and ungroomed spectra therefore helps separate the perturbative radiation pattern from soft wide-angle contributions and pileup/UE effects, while also probing the mapping between the AO evolution variable and angle and the interleaving prescription across QCD and EW sectors.

\begin{figure}[t]
\centering
\includegraphics[width=.49\textwidth]{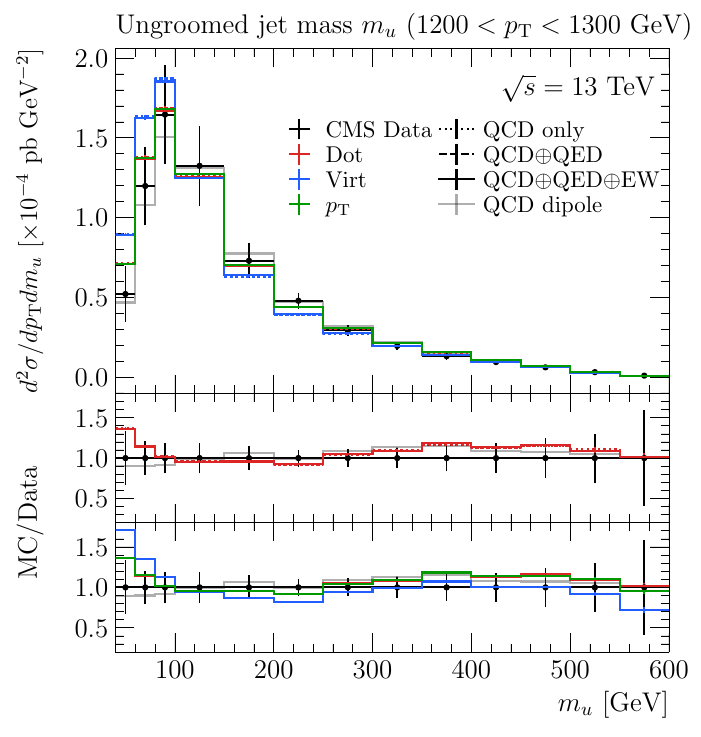}
\includegraphics[width=.49\textwidth]{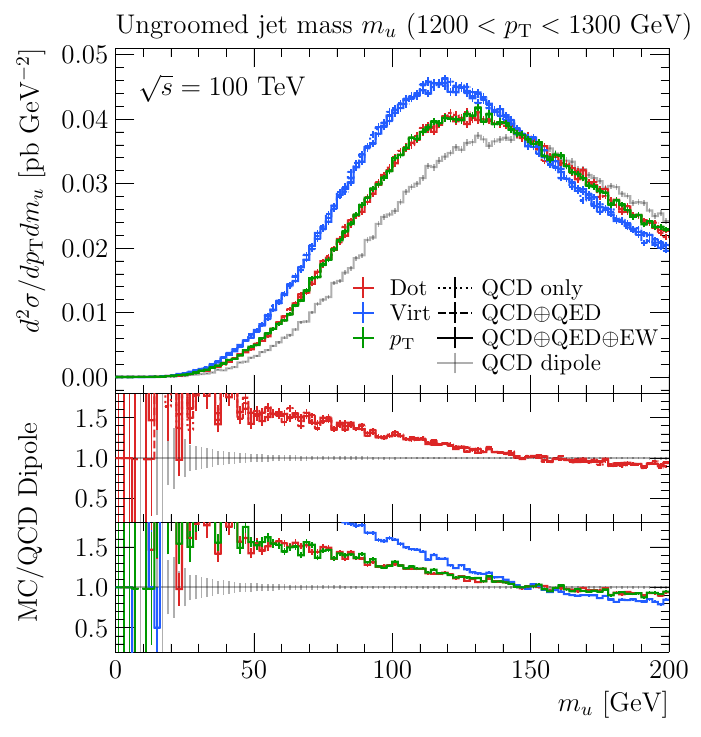}\\[4pt]
\includegraphics[width=.49\textwidth]{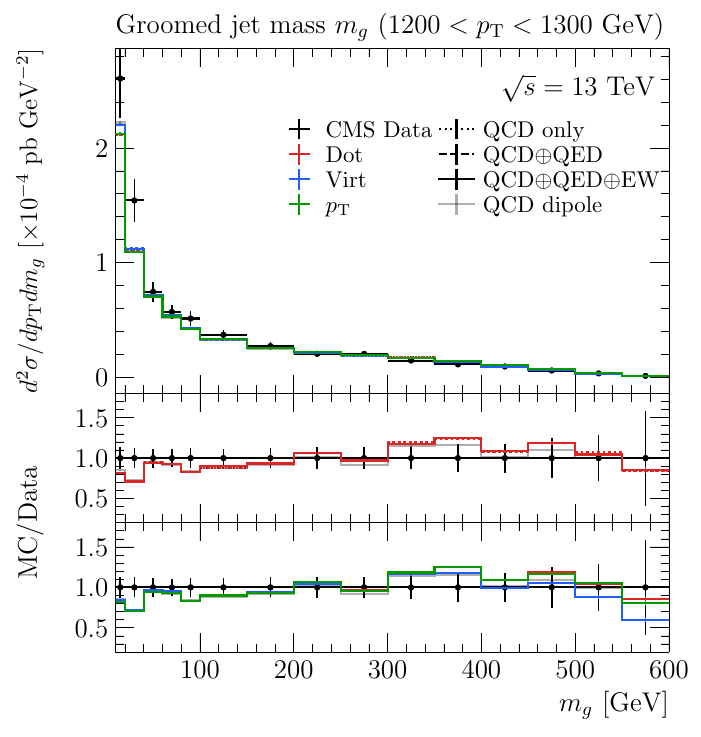}
\includegraphics[width=.49\textwidth]{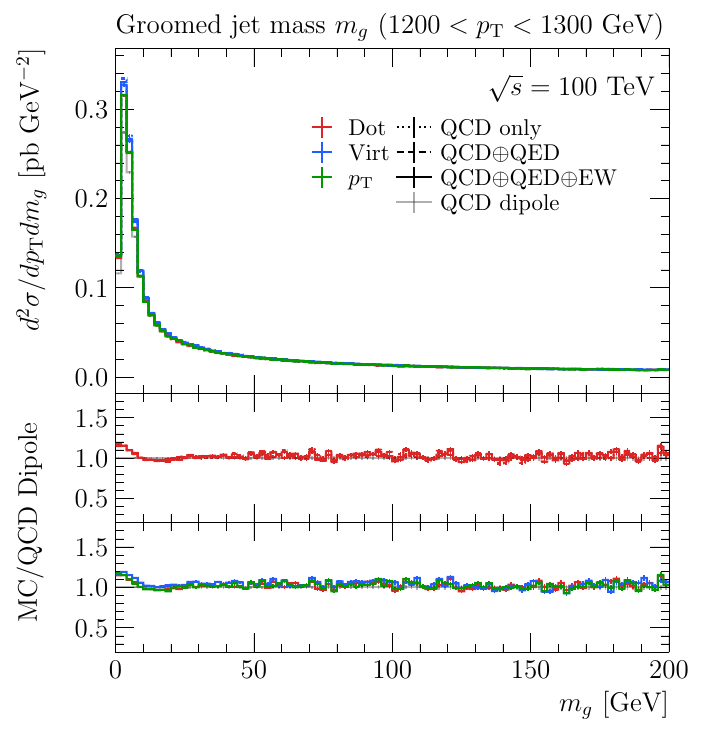}
\caption{Ungroomed (top row) and SoftDrop-groomed with $\beta=0$, $z_{\text{cut}}=0.1$ (bottom row) jet-mass spectra in dijet events for $1200<\pt<1300~\mathrm{GeV}$, with $ R=0.8$ for both ungroomed and groomed observables. Left panels show the CMS measurement at $\sqrt{s}=13~\mathrm{TeV}$~\cite{CMS:2018vzn}; right panels show particle-level predictions at $100~\mathrm{TeV}$. The notation follows that of figure~\ref{fig:ALEPH_2004_Thrust}.}
\label{fig:CMS_GroomedUngroomedMass}
\end{figure}

Figure~\ref{fig:CMS_GroomedUngroomedMass} complements the ATLAS mass study by contrasting ungroomed and groomed spectra with $\beta=0$, $z_{\text{cut}}=0.1$, and $ R=0.8$. The near-independence of the groomed distributions from the shower variant demonstrates the reduced sensitivity achieved through grooming, while differences confined to the ungroomed spectra would indicate sensitivity to recoil or to changes in wide-angle radiation.

\subsection{Jet Width (Girth)}

Jet width quantifies the transverse spread of momentum about the jet axis and is linearly sensitive to both soft transverse momentum and emission angle, providing a complementary probe of recoil and large-angle radiation. For an anti-$\kt$ jet $J$ with axis $\hat{\mathbf{n}}_J$ and constituents~$\{i\}$,
\begin{equation}
  w \;\equiv\; \frac{1}{p_{\text{T},J}}\sum_{i\in J} p_{\text{T},i}\,\Delta R_i
  \;=\; \sum_{i\in J} z_i\,\theta_i,
  \qquad 
  z_i \equiv \frac{p_{\text{T},i}}{p_{\text{T},J}},\;\; \theta_i \equiv \Delta R_i\, .
\end{equation}
In the soft-collinear limit for a single emission with $(z,\theta)\ll1$, $w\simeq z\theta$; emissions near the jet edge therefore contribute linearly. Jet width is consequently sensitive to changes in recoil and soft radiation, as well as to the mapping between the AO evolution variable and the emission angle.

\begin{figure}[t]
\centering
\includegraphics[width=.49\textwidth]{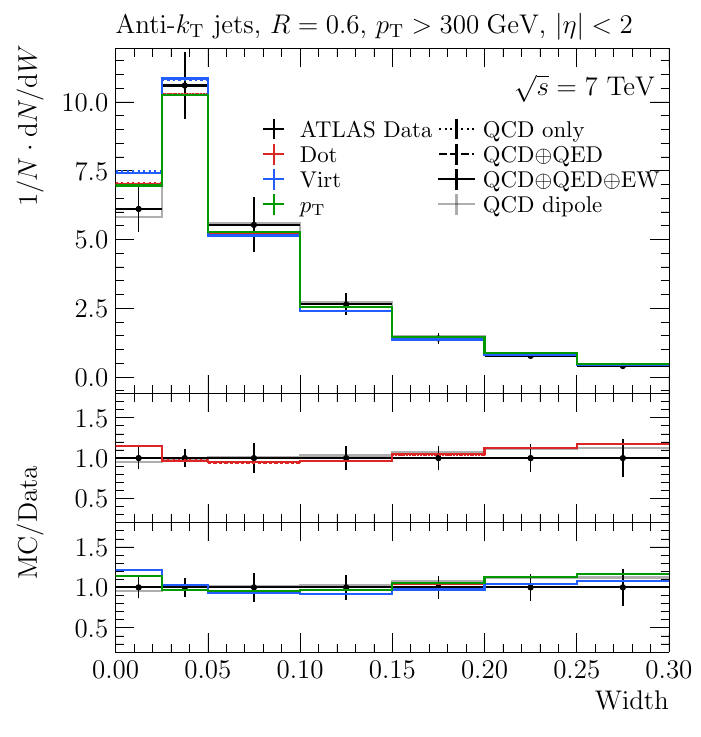}
\includegraphics[width=.49\textwidth]{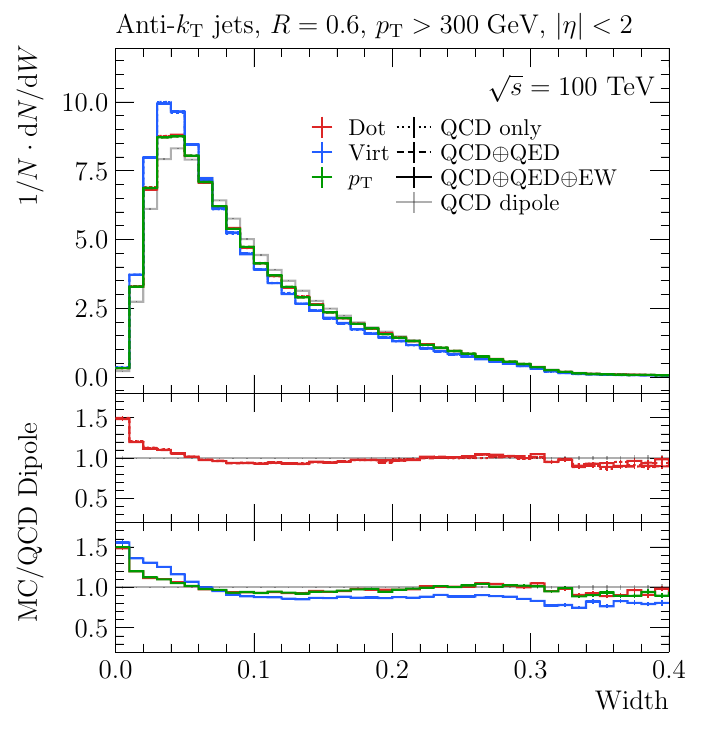}
\caption{Jet width (girth) for leading anti-$\kt$ jets at $\sqrt{s}=7~\mathrm{TeV}$ with $\pt>300~\mathrm{GeV}$ and $|\eta|<2$ (ATLAS, particle level)~\cite{ATLAS:2012nnf}, compared to particle-level predictions at $100~\mathrm{TeV}$. The notation follows that of figure~\ref{fig:ALEPH_2004_Thrust}.}
\label{fig:ATLAS_2012_Width}
\end{figure}

Figure~\ref{fig:ATLAS_2012_Width} presents the width observable. Stability of the small-$w$ Sudakov region tests whether the angular radiation pattern is retained under changes of kinematic reconstruction scheme and interleaving choice; hardening of the spectrum would signal an increase in wide-angle radiation. The linear angular weighting further sharpens sensitivity to recoil effects and to the angle-evolution mapping.

\subsection{Angularity Distributions of Leading Jets}
\label{subsec:atlas-angularities}

\begin{figure}[t]
\centering
\includegraphics[width=.49\textwidth]{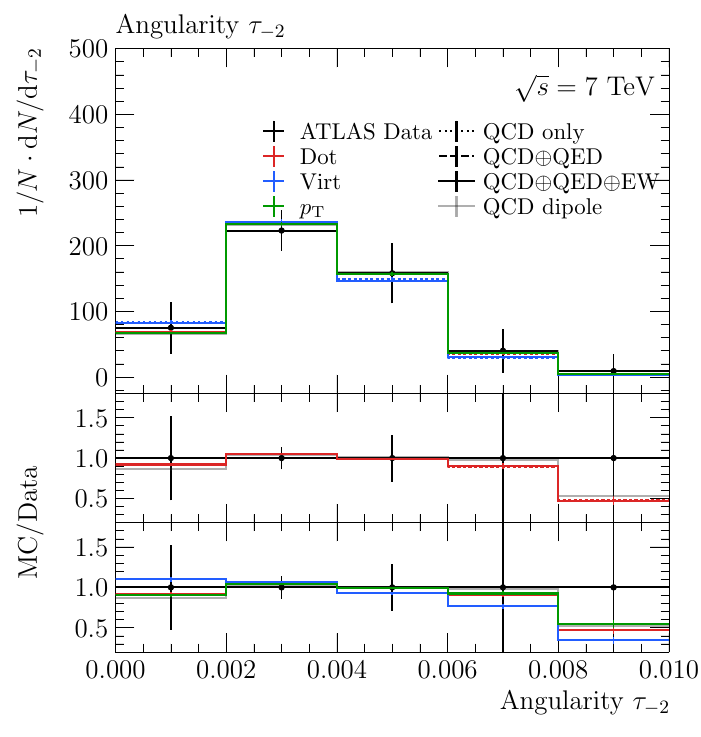}
\includegraphics[width=.49\textwidth]{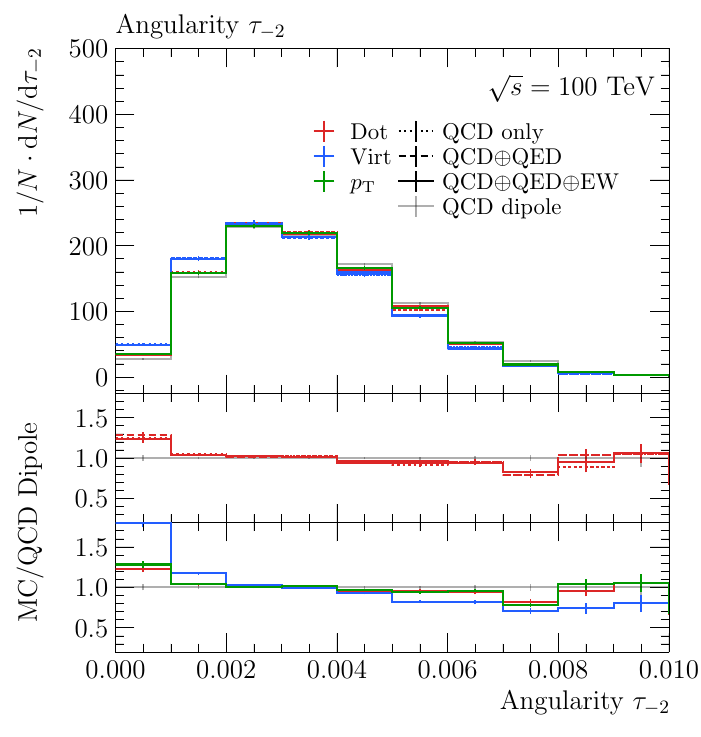}
\caption{Angularity $\tau_{-2}$ distributions of leading anti-$\kt$ jets at $\sqrt{s}=7~\mathrm{TeV}$ with $\pt>300~\mathrm{GeV}$, $|\eta|<0.7$, and $100<M<130~\mathrm{GeV}$ (ATLAS, particle level)~\cite{ATLAS:2012nnf}, compared to particle-level predictions at $100~\mathrm{TeV}$. The notation follows that of figure~\ref{fig:ALEPH_2004_Thrust}.}
\label{fig:ATLAS_2012_Angularities}
\end{figure}

Jet angularities form a family of observables with different angular weightings of radiation inside a jet. The measurement used in figure~\ref{fig:ATLAS_2012_Angularities} is expressed in terms of the ATLAS angularity $\tau_a$~\cite{ATLAS:2012nnf},
\begin{equation}
  \tau_a \;\equiv\;
  \frac{1}{m_J}
  \sum_{i\in J}
  E_i \sin^a\theta_i
  \left(1-\cos\theta_i\right)^{1-a}\, ,
\end{equation}
where $m_J$ is the jet mass and $\theta_i$ is the angle of constituent $i$ with respect to the jet axis. In the small-angle limit,
\begin{equation}
  \tau_a \;\simeq\; \frac{2^{a-1}}{m_J}
  \sum_{i\in J} E_i\,\theta_i^{2-a}\, .
\end{equation}
The observable shown below is $\tau_{-2}$, corresponding to $a=-2$. It is therefore not identical to the commonly used definition $\lambda_\beta=\sum_i z_i\theta_i^\beta$. However, in the soft-collinear limit it has an angular weighting proportional to $\theta^4$, similar to $\lambda_{\beta=4}$ up to the jet-mass normalisation and constant factors. The measured $\tau_{-2}$ spectrum therefore probes the angular distribution of radiation near the edge of the jet more strongly than any other variable we consider, as well as the mapping between the AO evolution variable and the emission angle.

Figure~\ref{fig:ATLAS_2012_Angularities} shows the $\tau_{-2}$ distribution. The same pattern seen in the other ungroomed leading-jet observables is visible here: the virtuality-preserving scheme gives a slightly lower prediction than the dot-product- and $\pt$-preserving schemes, while QED and EW emissions have no significant effect.

\subsection[Jet Shapes $\rho(r)$ and $\Psi(r)$]{\boldmath Jet shapes $\rho(r)$ and $\Psi(r)$}

\label{subsec:atlas-jet-shape}

Jet shapes resolve the radial flow of transverse momentum inside a jet, providing a differential handle on large-angle soft radiation, recoil, and interleaving effects. For an anti-$\kt$ jet $J$ of radius $R$ with axis $\hat{\mathbf{n}}_J$ and constituents $\{i\}$, define the integrated jet shape
\begin{equation}
  \Psi(r) \;\equiv\; \frac{1}{p_{\text{T},J}} \sum_{i\in J,\,\Delta R_i<r} p_{\text{T},i},
  \qquad 0 \le r \le R\, ,
\end{equation}
and the differential profile
\begin{equation}
  \rho(r) \equiv \frac{d\Psi(r)}{dr}
  \simeq
  \int \frac{\diffd\theta^2}{\theta^2}\, \diffd z\, P(z)
  \left[
    z\,\delta\!\left(r-(1-z)\theta\right)
    +(1-z)\,\delta\!\left(r-z\theta\right)
  \right] ,
  \quad \text{(single emission)}\, .
\end{equation}
Here $r$ denotes the rapidity-azimuth distance from the jet axis, with $0\le r\le R$~\cite{Seymour:1997kj}. Thus $\rho(r)$ resolves the radial distribution of transverse momentum and its recoil contribution, while $\Psi(r)$ accumulates the transverse momentum inside radius $r$. Since a small-angle splitting should not create an additional independent source of radiation at larger angles, the outer radial profile should remain stable as collinear substructure develops. Recoil artefacts or competition between shower sectors can instead populate the periphery and modify the approach of $\Psi(r)$ to unity. The jet-shape observables are therefore sensitive to the radial distribution of transverse momentum and to the mapping between the AO evolution variable and emission angle.

\begin{figure}[t]
\centering
\includegraphics[width=.49\textwidth]{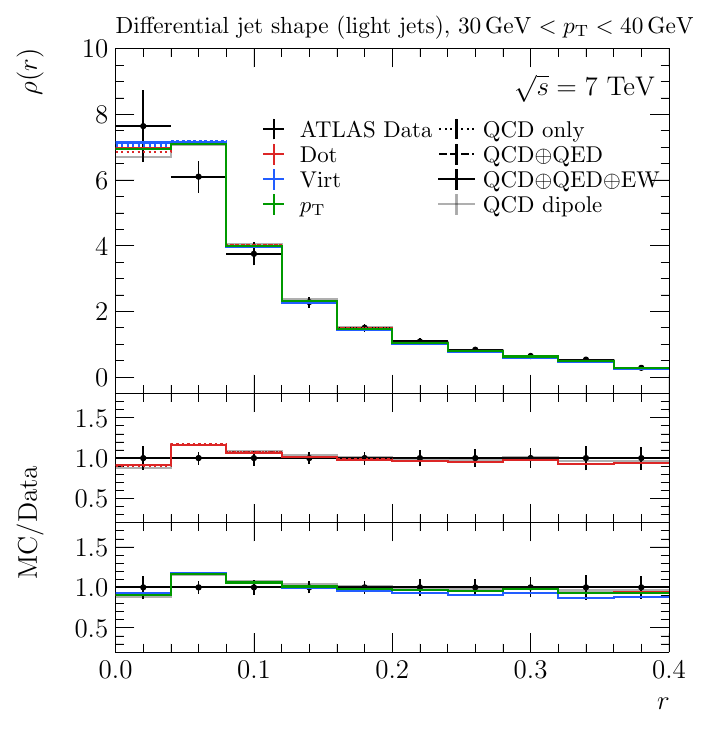}
\includegraphics[width=.49\textwidth]{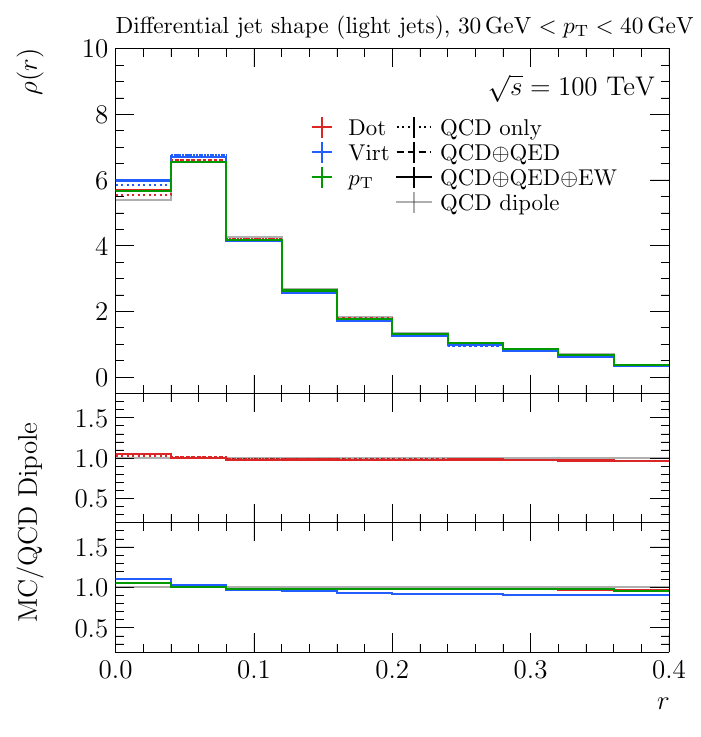}
\caption{Differential jet shape $\rho(r)$ for light jets in $t\bar t$ events with $30~\mathrm{GeV}<\pt<40~\mathrm{GeV}$ at $\sqrt{s}=7~\mathrm{TeV}$ (ATLAS, particle level)~\cite{ATLAS:2013uet}, compared to particle-level predictions at $100~\mathrm{TeV}$. The notation follows that of figure~\ref{fig:ALEPH_2004_Thrust}.}
\label{fig:ATLAS_2012_rho}
\end{figure}

\begin{figure}[t]
\centering
\includegraphics[width=.49\textwidth]{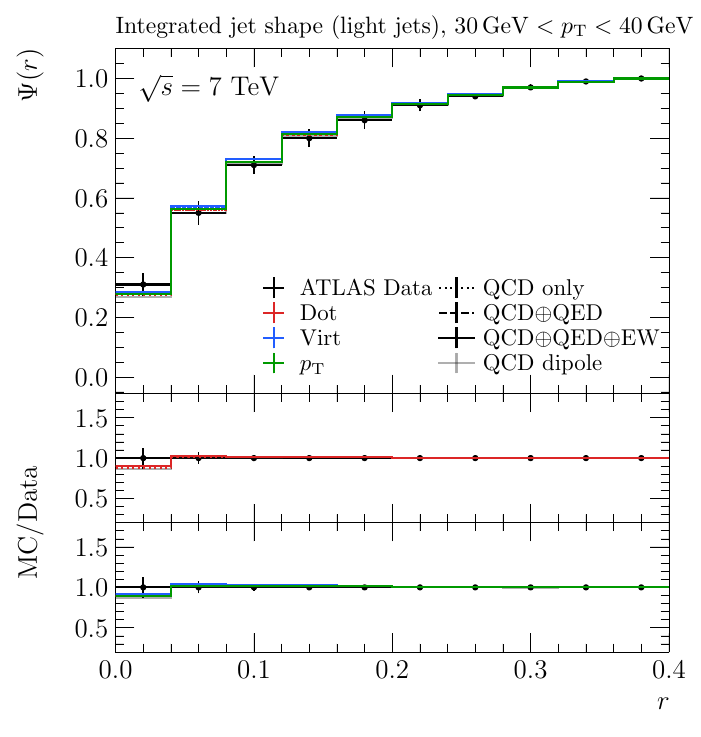}
\includegraphics[width=.49\textwidth]{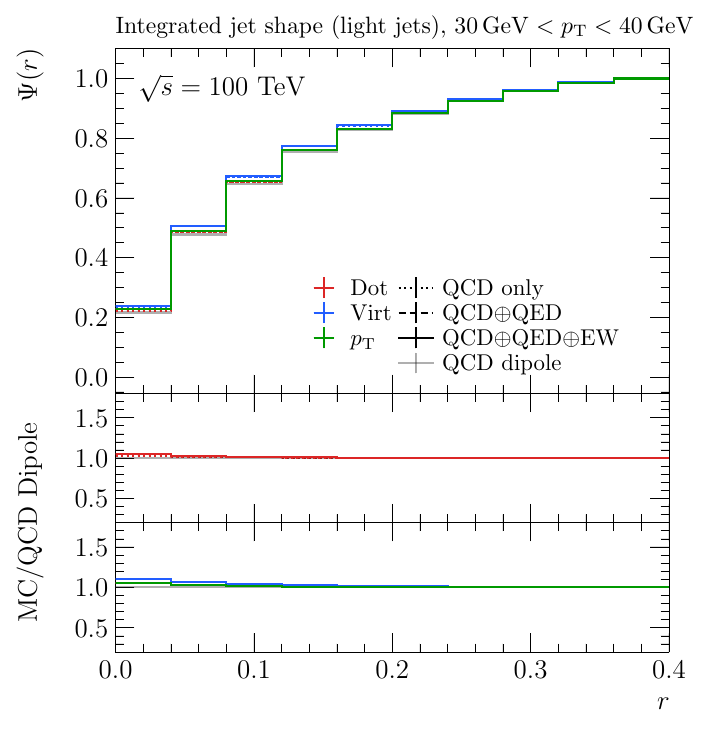}
\caption{Integrated jet shape $\Psi(r)$ for the same light-jet selection as figure~\ref{fig:ATLAS_2012_rho}. The notation follows that of figure~\ref{fig:ALEPH_2004_Thrust}.}
\label{fig:ATLAS_2012_Psi}
\end{figure}

Figures~\ref{fig:ATLAS_2012_rho} and~\ref{fig:ATLAS_2012_Psi} present the differential and integrated radial jet shapes. The differential profile follows the same pattern as the other ungroomed jet-structure observables: the virtuality-preserving prediction is slightly lower than the dot-product- and $\pt$-preserving predictions, with a somewhat larger separation at 100~TeV, while QED and EW emissions have no significant effect. Since $\Psi(r)$ is cumulative, it is less discriminating than $\rho(r)$, but it provides a useful cross-check of the overall radial energy containment. In this case, the integrated profile shows no significant differences among the shower variants.

Combining figures~\ref{fig:ATLAS_2012_JetMass}, \ref{fig:CMS_GroomedUngroomedMass} (upper row), \ref{fig:ATLAS_2012_Width}, \ref{fig:ATLAS_2012_Angularities}, \ref{fig:ATLAS_2012_rho} and~\ref{fig:ATLAS_2012_Psi}, the ungroomed jet-structure observables show a common pattern: the virtuality-preserving scheme tends to give slightly lower predictions than the dot-product- and $\pt$-preserving schemes, with a somewhat larger separation at 100~TeV, while QED and EW emissions have little visible impact. By contrast, the groomed jet-mass distributions in the lower row of figure~\ref{fig:CMS_GroomedUngroomedMass} show essentially no difference between the shower variants, demonstrating the reduced sensitivity achieved through grooming.

\subsection{Quantifying the Effect of Interleaving}
\label{subsec:coh-summary}

Across the observables, the visible spread is driven predominantly by the chosen AO-compatible kinematic reconstruction scheme rather than by the activation of QED or EW shower sectors. To quantify this statement, we define two bin-wise envelope measures. For a histogram $h$, bin $b$, kinematic reconstruction scheme $r\in\{\mathrm{Dot},\mathrm{Virt},\pt\}$, and shower-sector choice $s\in\{\qcdonly,\qcdqed,\qcdqedew\}$, let $T_{hbrs}$ denote the corresponding prediction. The interleaving-induced envelope is defined as
\begin{equation}
\Delta^{\rm int}_h =
\max_{r,b}
\left[
\frac{
\max_s T_{hbrs}-\min_s T_{hbrs}
}{
T_{hbr\,\mathrm{QCD}}
}
\right]\, ,
\label{eq:interleaving-envelope}
\end{equation}
while the scheme-induced envelope is defined as
\begin{equation}
\Delta^{\rm sch}_h =
\max_{s,b}
\left[
\frac{
\max_r T_{hbrs}-\min_r T_{hbrs}
}{
T_{hb\,\mathrm{Dot}\,s}
}
\right]\, .
\label{eq:scheme-envelope}
\end{equation}
The maxima are taken over the displayed bins at the measured centre-of-mass energies, excluding bins where the denominator is zero or where any of the three compared predictions has a relative statistical uncertainty above $2\%$, which we deem statistically significant. The resulting diagnostics are summarised in table~\ref{tab:envelopes}, where the envelopes are rounded up to the next per cent. They show that, for all observables considered in this scan, $\Delta^{\rm int}_h$ is smaller than $\Delta^{\rm sch}_h$. Thus, within the tested phase-space regions, QED/EW interleaving does not dominate the observed changes in the gluon radiation pattern; the leading numerical effect instead arises from the kinematic reconstruction scheme and recoil mapping.

\begin{table}[!t]
\centering
\begin{tabular}{|l|c|c|}
\hline
\hline\hline
Observable & $\Delta^{\rm int}_h$ & $\Delta^{\rm sch}_h$ \\
\hline
$e^+e^-$ thrust, $1-T$                & $\lesssim 4\%$ & $\lesssim 24\%$ \\
$C$-parameter                         & $\lesssim 6\%$ & $\lesssim 37\%$ \\
Total broadening, $B_T$               & $\lesssim 5\%$ & $\lesssim 28\%$ \\
Durham transition scale, $y_{23}$      & $\lesssim 7\%$ & $\lesssim 17\%$ \\
Leading-jet mass                      & $\lesssim 2\%$ & $\lesssim 15\%$ \\
Ungroomed jet mass                    & $\lesssim 3\%$ & $\lesssim 27\%$ \\
Groomed jet mass                      & $\lesssim 3\%$ & $\lesssim 12\%$ \\
Jet width                             & $\lesssim 1\%$ & $\lesssim 9\%$ \\
Leading-jet angularity, $\tau_{-2}$    & $\lesssim 2\%$ & $\lesssim 7\%$ \\
Differential jet shape, $\rho(r)$      & $\lesssim 2\%$ & $\lesssim 7\%$ \\
Integrated jet shape, $\Psi(r)$        & $\lesssim 2\%$ & $\lesssim 3\%$ \\
\hline
\end{tabular}
\caption{Envelope diagnostics for the observables used in the interleaving and kinematic reconstruction scheme scans. $\Delta^{\rm int}_h$ is the maximum bin-wise envelope obtained by changing the shower-sector choice at fixed kinematic reconstruction scheme, maximised over the kinematic reconstruction schemes. $\Delta^{\rm sch}_h$ is the maximum bin-wise envelope obtained by changing the kinematic reconstruction scheme at fixed shower-sector choice, maximised over the three sector choices. The quoted values are taken at the measured centre-of-mass energies.}
\label{tab:envelopes}
\end{table}

This does not imply that a blanket application of QCD angular ordering to the non-QCD sectors is optimal; it indicates the internal consistency of the present implementation for the observables and phase-space regions tested here. A more principled approach would derive sector-specific criteria from first principles and enforce them separately for each interaction, an avenue we leave for future work. Even in the QCD-only case, the description of data is not ideal, likely reflecting the current (v7.3) tune's emphasis on LEP final-state observables. This matters because LEP event shapes predominantly constrain quark-initiated final-state radiation, whereas the Tevatron and LHC jet observables used here contain a much larger gluon-jet component and are also sensitive to ISR, recoil against the beam system, and hadron-collider colour connections. A more balanced strategy should therefore tune simultaneously to LEP, Tevatron and LHC data, constraining both ISR and FSR while explicitly including bins from the resummation region in the fit. The remainder of this paper, therefore, turns to a dedicated tune along these lines.


\section{Tuning the Interleaved Shower with Cluster Hadronization}
\label{sec:tune}

After demonstrating the small envelope of interleaving emissions, we tune the $\qcdqedew$ shower together with the cluster hadronization model, using a simultaneous fit to a mix of lepton and hadron collider observables. The kinematic reconstruction scheme is not varied in this tune; it is fixed to the default dot-product-preserving prescription. Below we describe the parameters we tuned and the data we used, followed by our method and results.

\subsection{Parameters Tuned}

We vary 14 continuous parameters, each defined and explained below. These include shower parameters (ISR and FSR $\alphas(m_Z)$, the shower cutoff, and the intrinsic transverse-momentum width) and cluster hadronization parameters (fission thresholds, flavour selection weights, and hadron weights).  The CMW prescription, recoil map, smooth-veto settings, heavy-flavour masses, $g\to Q\bar Q$ thresholds, and MPI settings are kept fixed throughout this tune.

\subsubsection{Shower Parameters}

\begin{itemize}
  \item \texttt{AlphaQCDISR:AlphaIn}: boundary value of the strong coupling $\alphas$ used for initial-state radiation (ISR) in backward evolution. It normalises the ISR Sudakov form factor $\Delta_{\rm ISR}$, thereby affecting the hardness of initial-state branchings and, in turn, the boson/jet recoil. We use the default switches, which mean that this input value is defined to be in the CMW scheme, and the two-loop running with current quark mass thresholds is used.

  \item \texttt{AlphaQCDFSR:AlphaIn}: boundary value of $\alphas$ for final-state radiation (FSR) in the AO shower. It controls the overall hardness of intra-jet radiation and hence the slopes of Sudakov-sensitive tails (small $1{-}T$, small $C$, etc) and the small-mass region of jet-mass spectra; the same running as above applies.

  \item \texttt{IntrinsicPtGaussian}: root-mean-square width $\sigma_{\pt}$ of the primordial transverse momentum $\pt$ assigned to incoming partons prior to ISR. This smears colour-singlet transverse momentum at low scales and affects azimuthal decorrelations.

  \item \texttt{pTmin}: effective shower cutoff expressed in the transverse-momentum map $\pt(\tilde q,z)$ of the AO evolution variable $\tilde q$ (with momentum fraction $z$). Emissions with $\pt < \texttt{pTmin}$ are vetoed, defining the perturbative-to-hadronization handover. Increasing \texttt{pTmin} shortens the cascade, resulting in steeper Sudakov suppression, narrower jets, and lower soft multiplicities, while decreasing it enhances very-soft radiation.
\end{itemize}

\subsubsection{Hadronization Parameters}

The cluster hadronization model begins by forcing the gluons left at the end of the shower to split non-perturbatively into quark--antiquark pairs. Colour-connected quarks and antiquarks are then combined into colour-singlet clusters, which are regarded as highly excited hadron resonances. Most clusters decay directly into a hadron pair. A fraction of them, however, are too heavy for this picture to be reasonable, and these first undergo fission into two lighter clusters, which may themselves split again.

The kinematic threshold option for cluster fission was set to the current default, \texttt{Dynamic}. In this scheme, a light-flavour cluster of mass $M_{\rm cl}$ with constituent masses $m_{1,2}$ undergoes fission with probability
\begin{equation}
  P_{\rm fiss} = \left[ 1 + \left| \frac{T - \delta}{M_{\rm cl}^{\,a}} \right|^{\sigma} \right]^{-1} ,
  \qquad
  T = M_{\max}^{a} + (m_1 + m_2)^{a}\, ,
  \label{eq:pfiss}
\end{equation}
with masses in GeV, where $M_{\max}$, $a$, $\delta$ and $\sigma$ are tunable parameters defined below.
The criterion of the legacy (\texttt{Static}) cluster fission model, $M_{\rm cl}^{\,a} > T$, is recovered for $\delta \to 0$ and $\sigma \to \infty$. In a fission, the quark and antiquark inside the cluster separate, and a new quark--antiquark pair, each of constituent mass $m_q$, is drawn from the vacuum with flavour weights $\{\gamma_u, \gamma_d, \gamma_s\}$. The daughter-cluster masses are then sampled as
\begin{equation}
  M_i = m_i + (M_{\rm cl} - m_i - m_q)\,\mathcal{R}_i^{1/p} ,
  \qquad i = 1, 2\, ,
  \label{eq:fission-masses}
\end{equation}
where $p$ is a tunable parameter defined below and $\mathcal{R}_{1,2}$ are independent random numbers uniform on $(0,1)$. The pair is accepted only if
\begin{equation}
  M_i^2 > m_i^2 + m_q^2 + \delta_{\rm th} \quad (i = 1, 2) ,
  \qquad
  M_1^2 + M_2^2 < M_{\rm cl}^2\, ,
  \label{eq:fission-threshold}
\end{equation}
where $\delta_{\rm th}$ is a tunable parameter defined below; otherwise the flavour and masses are redrawn. Two-body kinematics additionally requires $M_1 + M_2 < M_{\rm cl}$. In the legacy model, eq.~\eqref{eq:fission-threshold} is replaced by the purely kinematic requirements $M_i > m_i + m_q$ and $M_1 + M_2 < M_{\rm cl}$.

\newpage

\noindent
The fission parameters we tune are:

\begin{itemize}
  \item $M_{\max} = \texttt{ClMaxLight}$: mass scale of the fission threshold $T$ in eq.~\eqref{eq:pfiss}, here for light quarks. Clusters much heavier than $M_{\max}$ almost always undergo fission, while lighter ones mostly decay directly. Larger values therefore allow heavier clusters to decay directly (fewer fissions), giving fewer and harder hadrons; smaller values increase the soft-hadron multiplicity.

  \item $a = \texttt{ClPowLight}$: power with which the masses enter eq.~\eqref{eq:pfiss}, here for light quarks. It sets how the constituent masses add to $M_{\max}$ in the threshold: the effective threshold mass $T^{1/a}$ falls from $M_{\max}+m_1+m_2$ at $a=1$ towards $M_{\max}$ as $a$ grows. Together with $M_{\max}$, it therefore shapes the cluster mass spectrum before decay, and with it the overall multiplicities and the softness of the final state.

  \item $\delta = \texttt{ProbabilityShift}$: shift of the threshold $T$ in eq.~\eqref{eq:pfiss}, in units of GeV$^{a}$ and common to all cluster flavours. It moves the cluster mass at which fission becomes likely: $P_{\rm fiss}=1/2$ for $M_{\rm cl}^{\,a} = |T-\delta|$. Positive values lower the effective threshold, giving more fissions; negative values raise it.

  \item $\sigma = \texttt{ProbabilityPowerFactor}$: exponent in eq.~\eqref{eq:pfiss}, common to all cluster flavours, which sets how sharply fission switches on as the cluster mass passes the threshold. Larger values sharpen the onset towards the step function of the legacy model, while smaller values smear it, so that some lighter clusters split and some heavier ones decay directly. It therefore shapes the tail of the cluster mass spectrum and with it the soft-hadron spectra. 

  \item $p = \texttt{PSplitLight}$: exponent in eq.~\eqref{eq:fission-masses}, here for light quarks, so that the daughter masses are sampled with density $\propto (M_i-m_i)^{p-1}$. For $p=1$ the distribution is flat; values above one favour heavy daughter clusters, which tend to split again, while values below one favour light ones. This feeds directly into soft spectra and large-$r$ jet-shape tails.

  \item $\delta_{\rm th} = \texttt{KineticThresholdShift}$: shift, in GeV$^2$, of the lower bound on the daughter-cluster masses in eq.~\eqref{eq:fission-threshold}, common to all cluster flavours. Since the daughter masses are always drawn above $m_i+m_q$, the bound only acts for $\delta_{\rm th} > 2m_im_q$, i.e.\ above about $0.21~\text{GeV}^2$ for light quarks. Larger values then force heavier daughter clusters, which changes the cluster mass spectrum after fission and hence the low-momentum hadron spectra.

  \item $\gamma_s = \texttt{PwtSquark}$: relative weight for non-perturbative $s\bar s$ pair creation, i.e.\ the strangeness-suppression factor (with $\gamma_s<1$ indicating suppression relative to $u,d$). It enters both cluster fission and cluster decays; the corresponding $u$ and $d$ weights, $\gamma_u = \texttt{PwtUquark}$ and $\gamma_d = \texttt{PwtDquark}$, are fixed to 1, and the separate weight for $s\bar s$ in gluon splitting was not tuned. It controls global strangeness yields ($K$, $\phi$, hyperons) and strange-hadron fractions in jets.
\end{itemize}
The remaining parameters act only in cluster decays, and we tune them as well:
\begin{itemize}
  \item \texttt{PwtDIquark}: weight for producing a diquark--antidiquark pair $[qq][\bar q\bar q]$ rather than $q\bar q$ when a cluster decays. A cluster heavier than the lightest baryon pair of its flavour decays to a baryon and an antibaryon with probability $\texttt{PwtDIquark}/(1+\texttt{PwtDIquark})$, and to mesons otherwise; lighter clusters always decay to mesons. Diquarks are not produced in cluster fission. Increasing \texttt{PwtDIquark} raises the baryon-to-meson ratio $B/M$ and the rates of octet/decuplet baryons; this affects identified-baryon spectra and jet baryon content.

  \item \texttt{SngWt}: weight for singlet ($\Lambda$-like) baryons in cluster decays. The weight of each such baryon, $\Lambda$ and its heavy-flavour analogues ($\Lambda_c$, $\Xi_c$, $\Lambda_b$, $\Xi_b$, \dots), is multiplied by $\texttt{SngWt}^2$, relative to unity for the other spin-$1/2$ baryons. It therefore controls the rates of these states relative to the other spin-$1/2$ baryons, most directly the $\Lambda$ yield and the $\Lambda/\Sigma$ and $\Lambda/p$ ratios.

  \item \texttt{DecWt}: weight for decuplet baryons in cluster decays. The weight of each spin-$3/2$ baryon is multiplied by $\texttt{DecWt}^2$, relative to unity for the spin-$1/2$ baryons (and $\texttt{SngWt}^2$ for the $\Lambda$-like ones). It directly modulates the rates of spin-$3/2$ decuplet states ($\Delta$, $\Sigma^*$, $\Xi^*$, $\Omega$) relative to spin-$1/2$ octet states ($N$, $\Lambda$, $\Sigma$, $\Xi$), thereby controlling high-mass baryon production and subsequent feed-down patterns.
\end{itemize}

\subsection{Observables Used for Tuning}

We tune to the observables from the analyses listed in section~\ref{sec:simu} (189 histograms from ALEPH, 37 from CDF, 144 from ATLAS and 94 from CMS). In addition, we add three more analyses for $e^+e^-$ at the $Z$ pole:
\begin{itemize}
  \item L3 (\href{https://rivet.hepforge.org/analyses/L3_2004_I652683.html}{\texttt{L3\_2004\_I652683}}) flavour-tagged event shapes by L3~\cite{L3:2004cdh}. These include thrust, the $C$-parameter, total and wide jet broadening, and the Durham $y_{23}$ transition scale for $dusc$ and $b$-quark events (10 histograms).
  \item OPAL (\href{https://rivet.hepforge.org/analyses/OPAL_2004_I648738.html}{\texttt{OPAL\_2004\_I648738}}) fragmentation functions of non-$b$-quark jets by OPAL~\cite{OPAL:2004prv}, complementing the ATLAS $t\bar t$ data (1 histogram).
  \item PDG (\href{https://rivet.hepforge.org/analyses/PDG_HADRON_MULTIPLICITIES.html}{\texttt{PDG\_HADRON\_MULTIPLICITIES}}) identified hadron rates from the PDG tables~\cite{ParticleDataGroup:2024cfk} (32 data points).
\end{itemize}

For the weight factors of these analyses, we give most of the hadron collider data, as well as the PDG-identified particle rates, a weight\footnote{The absolute weight does not affect the fit, only the relative weights of different analyses. Thus, the choice to make the minimum weight 10 is entirely arbitrary.} of 100. The ALEPH analysis contains the same event-shape and jet-rate distributions at multiple energies, and is therefore assigned a lower weight of 50 to prevent it from dominating the tune by its number of bins alone. The CMS analysis of groomed and ungroomed jet-mass spectra in $V{+}$jets and dijets is also assigned a weight of 50; it provides an important jet-structure constraint but is treated as an intermediate-weight input rather than a primary hadron-collider constraint. All L3 observables, and most additional hadronization and fragmentation constraints, are assigned a baseline weight of 10, including light-quark fragmentation.

\subsection{Tuning Method}

We used the reduced $\chi^2$ as our goodness-of-fit measure. When weights are involved, the definitions of $\chi^2$ and reduced $\chi^2$ must be modified to account for per-histogram weights. Defining $\mathcal{H}$ as the set of histograms included in the fit, with weight $w_h$ for $h\in\mathcal{H}$, and $\mathcal{B}_h$ as the set of bins of histogram $h$ entering the objective, and  $(y_{hb},\sigma_{hb})$ as the data value and its quoted uncorrelated uncertainty, the $\chi^2$ and reduced $\chi^2$ are defined as
\begin{equation}
  \chi^2(\vec{\theta})
  =
  \sum_{h\in\mathcal{H}}
  w_h
  \sum_{b\in\mathcal{B}_h}
  \frac{\left(y_{hb}-T_{hb}(\vec{\theta})\right)^2}{\sigma^2_{hb}},
  \qquad
  \chi^2_{\rm red}
  =
  \chi^2/
  \sum_{h\in\mathcal{H}}
  w_h
  |\mathcal{B}_h|\, .
  \label{eq:chisq-weights}
\end{equation}

\begin{figure}[ht]
\centering
\includegraphics[width=0.8\linewidth]{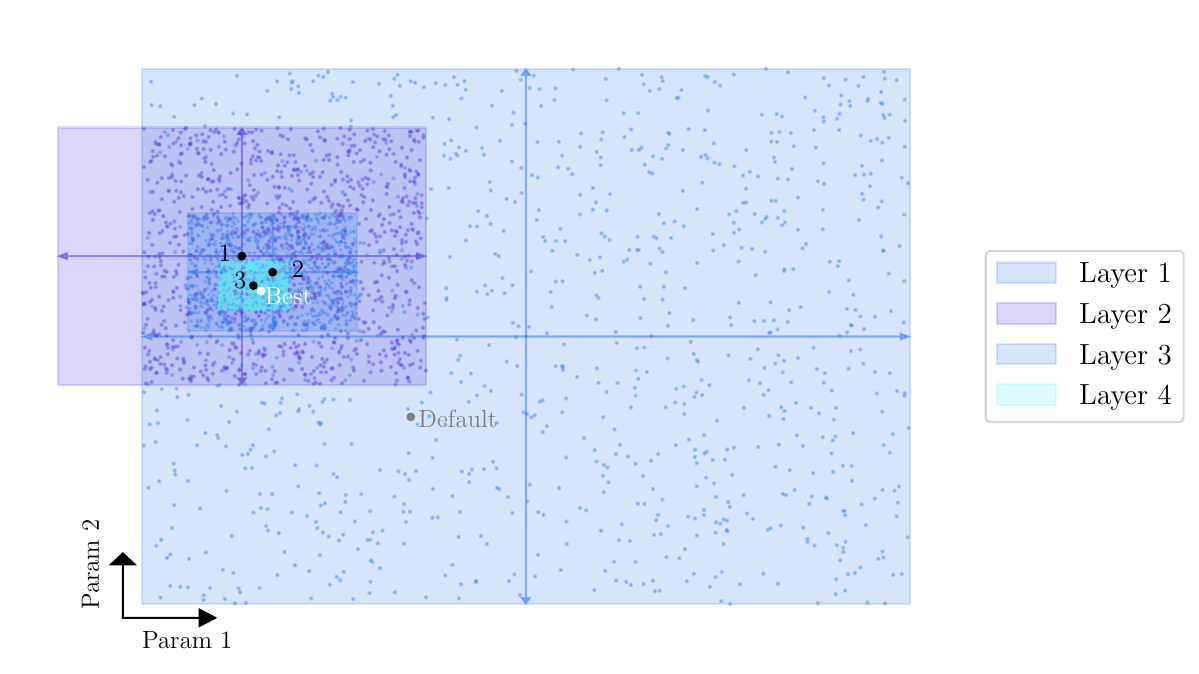}
\caption{Illustration of the tuning approach for a lower-dimensional example. Consider two parameters $x$ and $y$ that have values between $(x_0,y_0)$ and $(x_1,y_1)$. The tune for the previous model (which we also define as ``Default'') lies within this region. After generating samples in this space, the sample with the lowest $\chi^2$ (1) is found. We then generate a rectangular space of half the size of the original rectangle, centred around (1). When generating samples, we ensure they are not produced outside the first rectangle, where they would lie outside the allowed $x, y$ values. After this, (2) and (3) are subsequently found, and lastly, the new tune or the ``Best'' is found. In this way, the area of interest is progressively zoomed into.}
\label{fig:tune-diagram}
\end{figure}

The systematic approach to tuning event generators is through polynomial emulation of the generator response to parameter variations, as implemented in \textsc{Professor}~II~\cite{Buckley:2009bj}. However, when tuning 14 parameters, we observed parameter-direction degeneracies and non-physical emulator extrapolations in sparsely sampled corners. For the next-best approach, we followed the Herwig~7.3 workflow, adapted to the current shower options and observable set~\cite{Masouminia:2023zhb}. Here, we generated ensembles of MC points from a bounded hypercube around the nominal settings using Latin-hypercube sampling with hard parameter bounds and physicality vetoes. The scan used four layers, with $N_{\rm scan}=1500$ points per layer; after each layer the hypercube was recentred on the current minimum, and all parameter half-ranges were reduced by a factor of $2$. We generated each scan point with $N_{\rm evt}=250,000$ events per run, using statistically independent random seeds. For the hadron collider observables, multiple runs with different hard-process $\pt$ ranges were combined to give full coverage. We computed the global $\chi^2$ with \texttt{prof2-chisq} using the published bin-by-bin uncertainties together with a theory floor and per-histogram weights,\footnote{\textsc{Professor} versions since 2.2.2 have interpreted input weights as acting on individual pulls, equivalent to squaring the weight in our formula eq.~\eqref{eq:chisq-weights}. However, we have implemented eq.~\eqref{eq:chisq-weights} as written.} and selected the minimum-$\chi^2$ point. We rejected points if the Herwig run failed, violated the hard bounds on the parameters, or produced empty/ill-defined histograms entering the objective. To prevent over-constraining ultra-precise bins, we applied a $f_{\rm floor}= 5\%$ theory uncertainty floor (the so-called ``Webber floor'') per bin~\cite{Reichelt:2017hts},
\begin{equation}
  \sigma_{hb}^2
  \to
  \max\!\left(\sigma_{hb},\, f_{\rm floor} \cdot y_{hb}\right)^2\, .
\end{equation}
Correlated systematics were not profiled, and no additional regularisation of nuisance shifts was applied. A schematic of the iterate-recentre-shrink loop is shown in figure~\ref{fig:tune-diagram}.

\newpage

\subsection{Tuning Results}

\begin{table}[!t]
\centering
\begin{tabular}{|l|D{.}{.}{3.4}|D{.}{.}{3.4}|c|}
\hline
\multicolumn{4}{c}{\textbf{Parton shower parameters}}\\
\hline\hline
Parameter & \multicolumn{1}{c|}{Herwig~7.3} & \multicolumn{1}{c|}{Tuned} & Scan Range \\
\hline
\texttt{AlphaQCDFSR:AlphaIn}                     & 0.102 & 0.113 & $[0.10,\,0.13]$ \\
\texttt{AlphaQCDISR:AlphaIn}                     & 0.119 & 0.114 & $[0.10,\,0.13]$ \\
\texttt{pTmin} (GeV)                             & 0.655 & 0.957 & $[0.8,1.2]$ \\
\texttt{IntrinsicPtGaussian} (GeV)               & 2.20  & 2.17  & $[0.0,5.0]$ \\
\hline\hline
\multicolumn{4}{c}{\textbf{Cluster hadronization parameters}}\\
\hline\hline
\texttt{ClMaxLight} (GeV)                       & 3.53     & 2.83     & $[2.5,\,4.0]$ \\
\texttt{ClPowLight}                             & 1.85     & 2.47     & $[0.5,\,3.0]$ \\
\texttt{PSplitLight}                            & 0.914    & 0.877    & $[0.5,\,2.0]$ \\
\texttt{ProbabilityPowerFactor}                 & 6.49     & 5.98     & $[1.0,\,10.0]$ \\
\texttt{ProbabilityShift}                       & -0.879 & -0.396 & $[-10.0,\,10.0]$ \\
\texttt{KineticThresholdShift} (GeV$^2$)        & 0.0884   & 0.260    & $[-10.0,\,10.0]$ \\
\texttt{PwtSquark}                              & 0.374    & 0.303    & $[0.1,\,1.5]$ \\
\texttt{PwtDIquark}                             & 0.331    & 0.634    & $[0.1,\,1.5]$ \\
\texttt{SngWt}                                  & 0.891    & 0.359    & $[0.0,\,1.0]$ \\
\texttt{DecWt}                                  & 0.416    & 0.382    & $[0.0,\,1.0]$ \\
\hline\hline
$\chi^2_{\rm red}$ ($N_{\rm bin}=6067$)         & 5.05     & 3.22     & -- \\
\hline
\end{tabular}
\caption{Herwig~7.3 release values and final tuned values for the parton shower and cluster hadronization parameters in the fixed $\dotqcdqedew$ setup. The $\chi^2_{\rm red}$ values compare the Herwig~7.3 and tuned runs. The scan presents substantial changes in the parameter values. Among these, the changes of interest include FSR and ISR $\alphas$ becoming much closer, \texttt{pTmin} increasing, and \texttt{ClMaxLight} decreasing. A different study~\cite{Holguin:2026trd} had observed correlation between these three parameters, after computing 2D heatmaps of $\chi^2$ as a function of pairs of these three parameters.}
\label{tab:tune}
\end{table}

The tuned parameter values are collected in table~\ref{tab:tune}. The values described as Herwig~7.3 there, and used in section~\ref{sec:simu}, are the default values of the previous version of Herwig, which were tuned to a different (but overlapping) dataset to the one discussed here. Thus, it might be expected to give a fair starting description of the data. Using \texttt{prof2-chisq} and the observables described earlier, we find that relative to this default set, the tune reduces the $\chi^2_{\rm red}$ by about $36\%$, from $5.05$ to $3.22$, for $N_{\rm bin}=6067$ bins. We use this reduction as an optimisation diagnostic rather than a statistical goodness-of-fit probability, since the fit combines heterogeneous LEP, Tevatron, and LHC measurements, includes a $5\%$ theory floor, and does not profile correlated experimental systematics. A non-trivial consistency check is provided by the two shower coupling parameters in table~\ref{tab:tune}. Although the ISR and FSR boundary values of $\alphas$ were allowed to float independently, the tune returns very similar values, $\alphas^{\rm ISR}={0.114}$ and $\alphas^{\rm FSR}={0.113}$. It is reassuring that the mixed LEP, Tevatron, and LHC fit does not drive the initial- and final-state shower couplings in incompatible directions, unlike previous studies that tuned $\alphas^{\rm FSR}$ to LEP data and then $\alphas^{\rm ISR}$ to LHC data. It therefore supports the consistency of using a common perturbative shower framework for both initial- and final-state radiation in the present tune. Note, however, that these values are in the CMW scheme, corresponding to an $\overline{\mathrm{MS}}$ value of around $0.107$, smaller than the world average.

\newpage

\begin{figure}[!t]
\centering
\includegraphics[width=.32\textwidth]{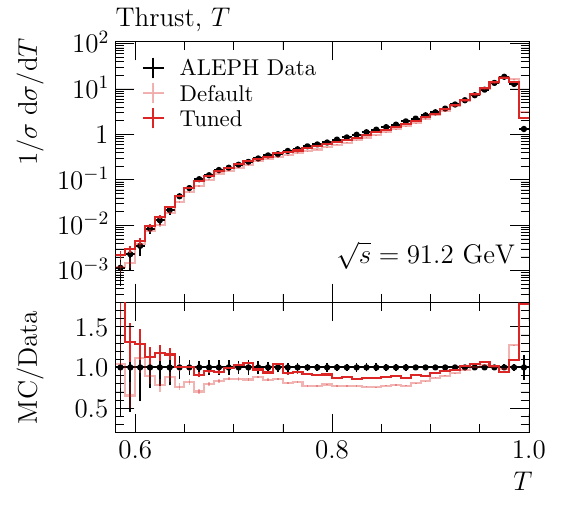}
\includegraphics[width=.32\textwidth]{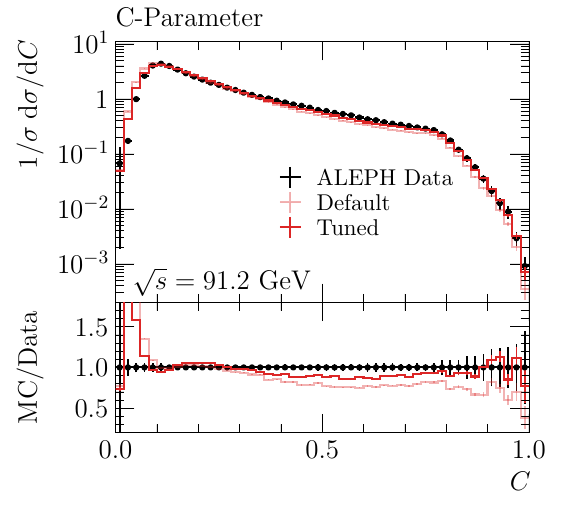}
\includegraphics[width=.32\textwidth]{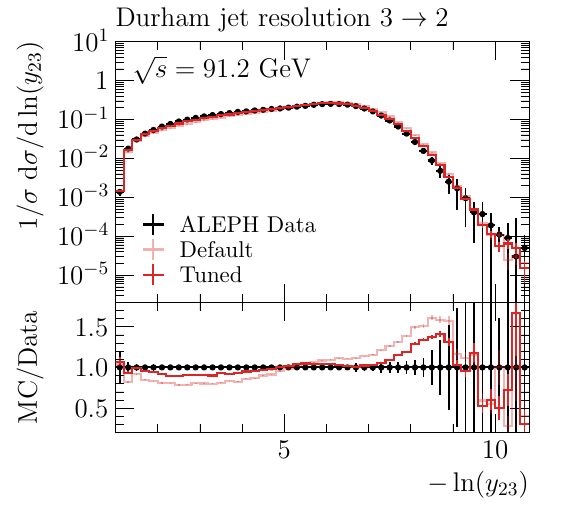}
\includegraphics[width=.32\textwidth]{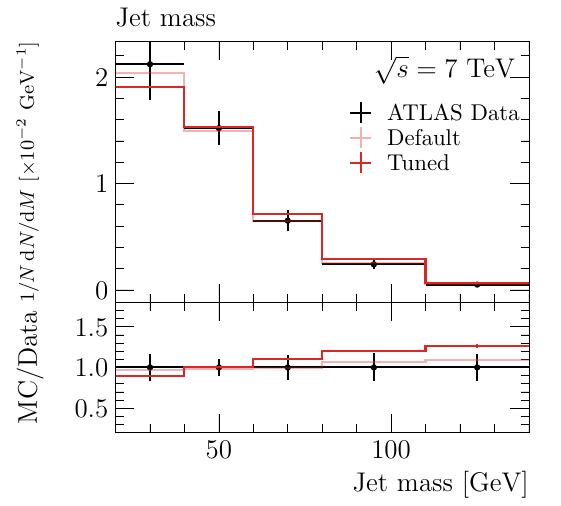}
\includegraphics[width=.32\textwidth]{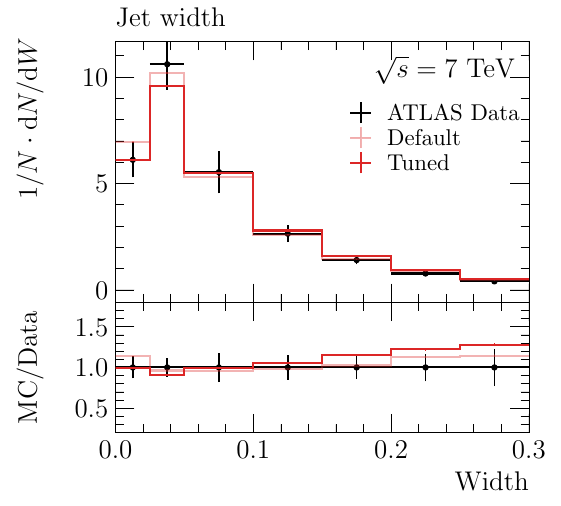}
\includegraphics[width=.32\textwidth]{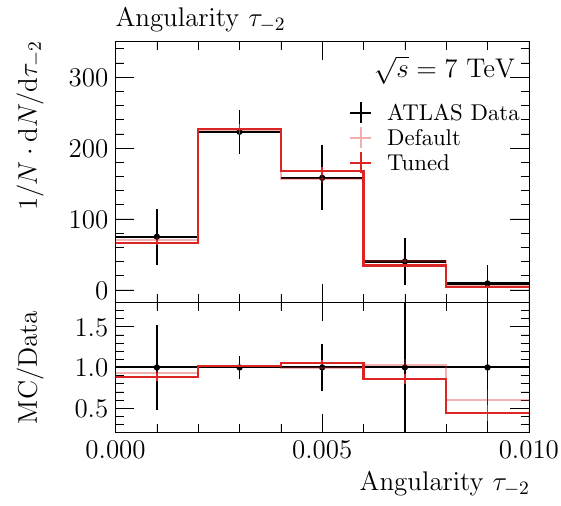}
\caption{Results using Herwig~7.3 and tuned parameters, for plots from section~\ref{sec:simu}. The data are taken from ALEPH~\cite{ALEPH:2003obs} and ATLAS~\cite{ATLAS:2012nnf}. Although the ATLAS data show a slight deterioration in the distributions, the ALEPH results show a significant improvement, attributed to the increase in the final state $\alphas$.}
\label{fig:tune}
\end{figure}

\begin{figure}[!t]
\centering
\includegraphics[width=.32\textwidth]{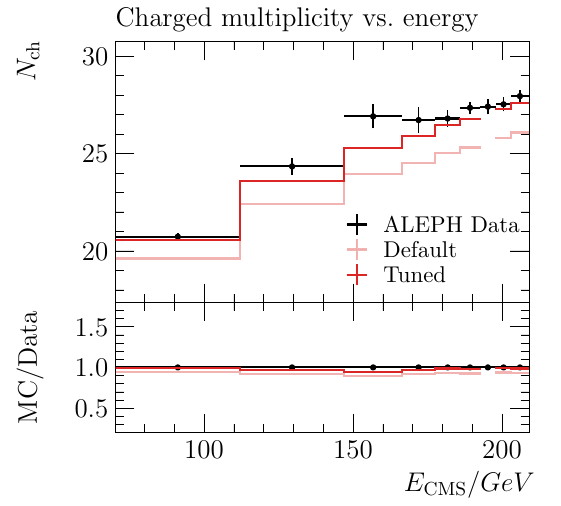}
\includegraphics[width=.32\textwidth]{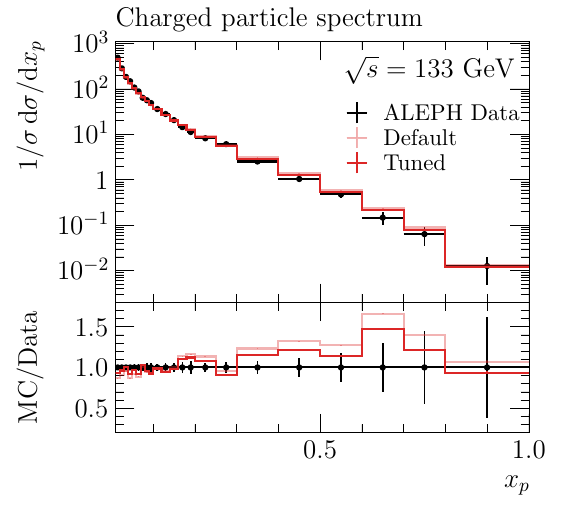}
\includegraphics[width=.32\textwidth]{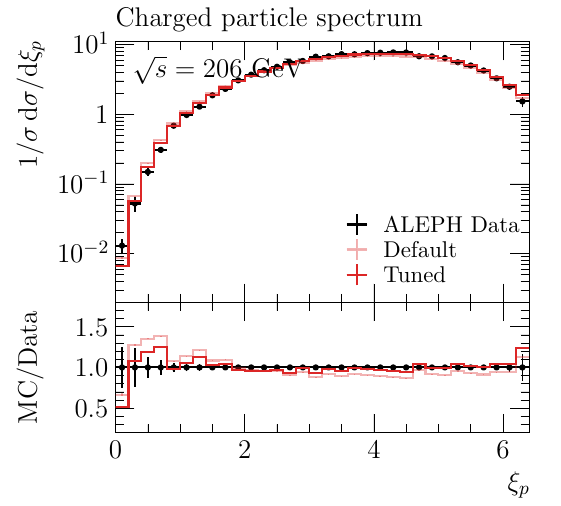}
\includegraphics[width=.32\textwidth]{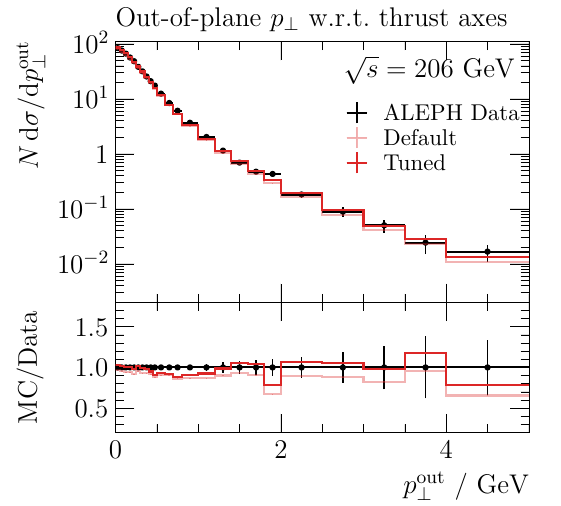}
\includegraphics[width=.32\textwidth]{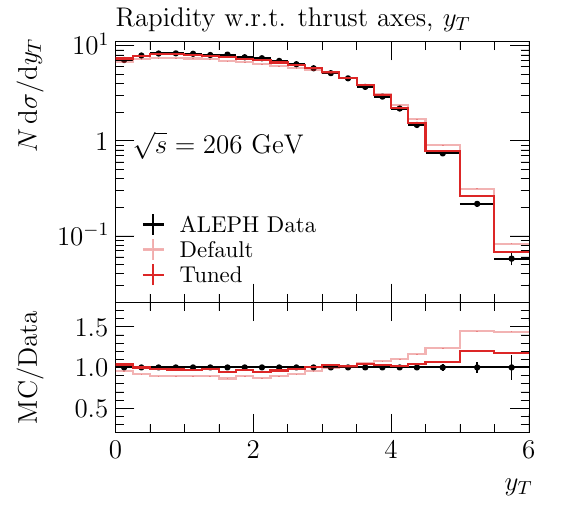}
\includegraphics[width=.32\textwidth]{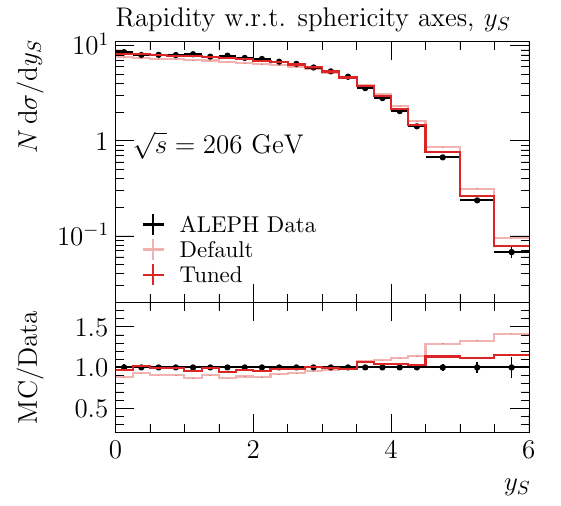}
\caption{$e^+e^-$ hadronization benchmarks. The data are taken from ALEPH~\cite{ALEPH:2003obs}. We see a significant improvement in the charged multiplicity and a mild improvement in all other distributions.}
\label{fig:other-obs}
\end{figure}

\begin{figure}[!t]
\centering
\includegraphics[width=.32\textwidth]{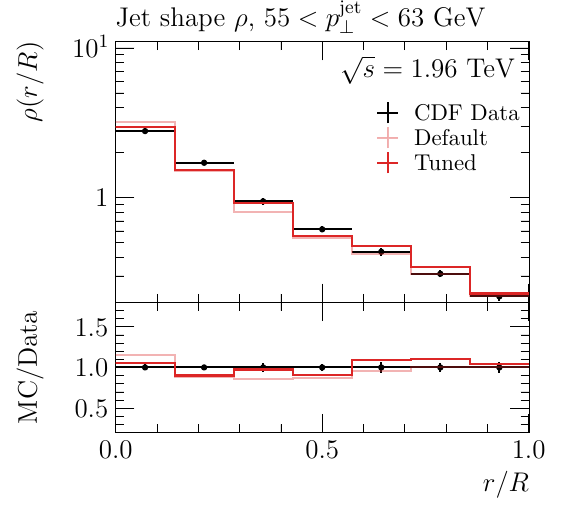}
\includegraphics[width=.32\textwidth]{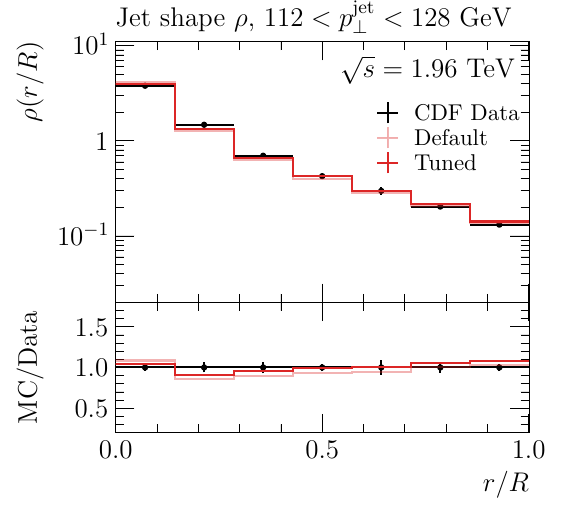}
\includegraphics[width=.32\textwidth]{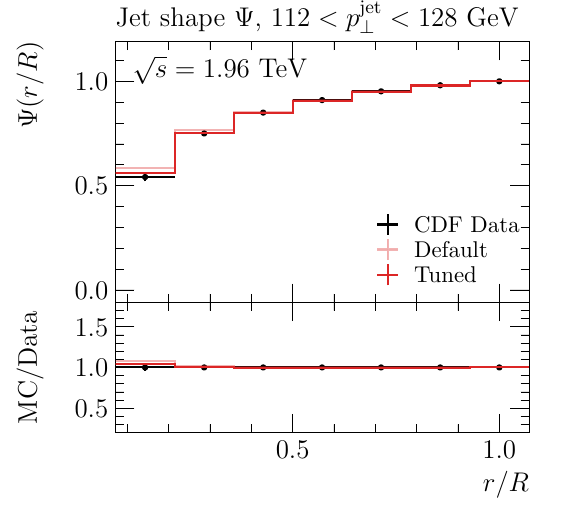}
\includegraphics[width=.32\textwidth]{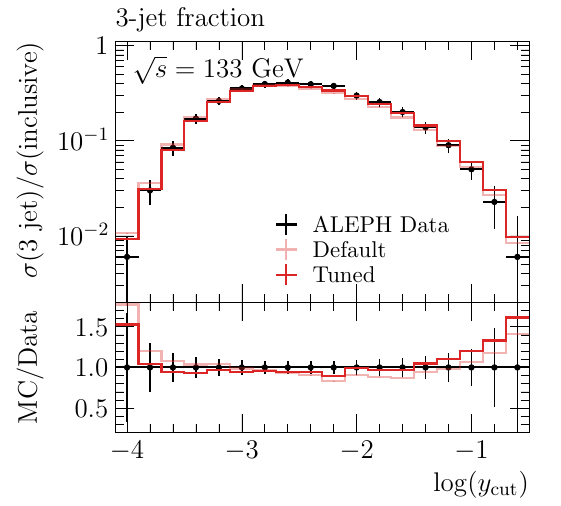}
\includegraphics[width=.32\textwidth]{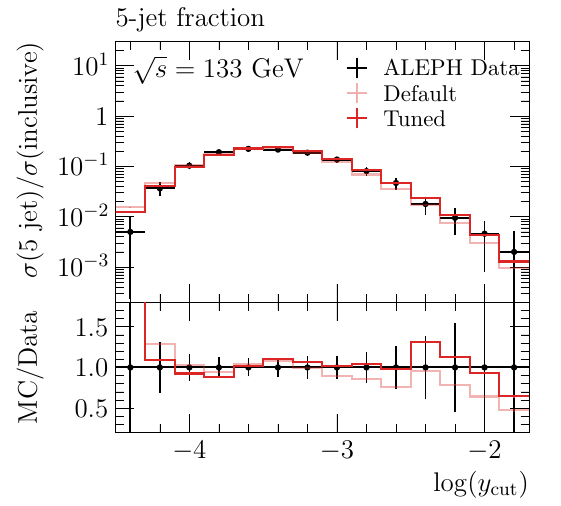}
\includegraphics[width=.32\textwidth]{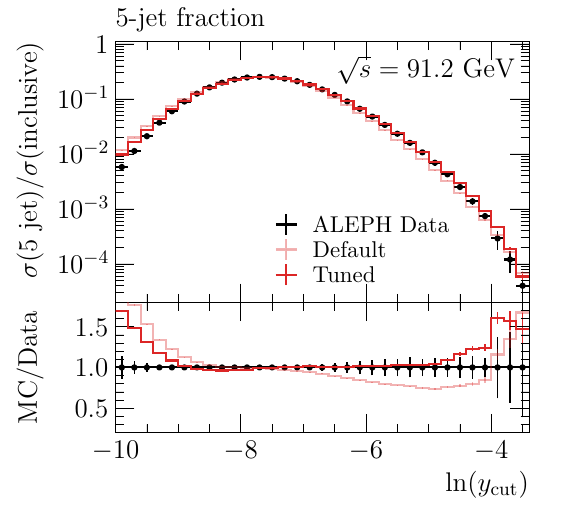}
\includegraphics[width=.32\textwidth]{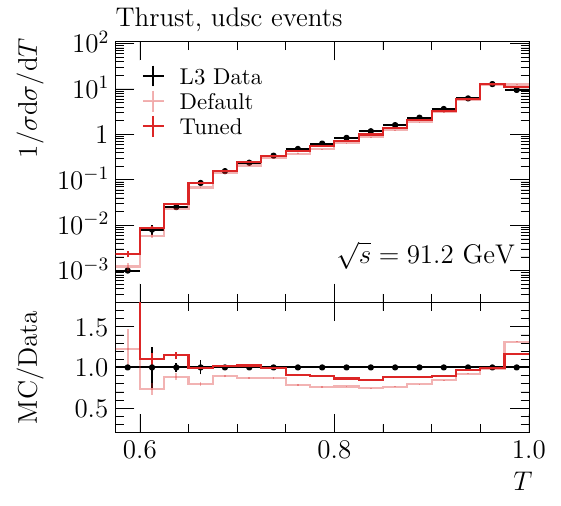}
\includegraphics[width=.32\textwidth]{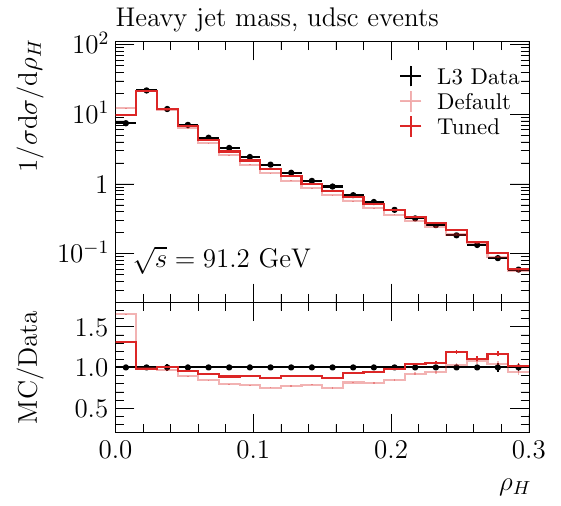}
\includegraphics[width=.32\textwidth]{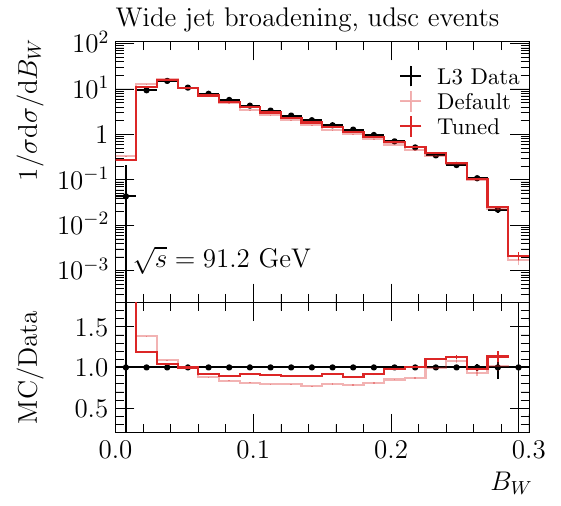}
\caption{Additional jet and event shape benchmarks. The data are taken from CDF~\cite{CDF:2005prv}, ALEPH~\cite{ALEPH:2003obs} and L3~\cite{L3:2004cdh}.}
\label{fig:more-obs}
\end{figure}

For comparison and validation, figure~\ref{fig:tune} overlays the default and tuned predictions for some of the analyses enumerated in section~\ref{sec:simu}. The improvement comes from better control of wide-angle soft radiation in event and jet shapes, along with an improved description of inclusive charged-particle multiplicities and spectra, while keeping the LHC jet-substructure description broadly stable, with some local deterioration in the jet-mass and width distributions. The post-tune predictions display stable MC/Data ratios in the resummation regions of the event shapes and jet rates. Figure~\ref{fig:other-obs} shows other, more hadron-oriented ALEPH charged-particle measurements across energies, while figure~\ref{fig:more-obs} adds CDF jet-shape observables, ALEPH jet fractions, and L3 flavour-separated event-shape distributions. 

The quantities shown in figure~\ref{fig:other-obs} are:
(i) the charged multiplicity $N_{\rm ch}(Q)$, i.e.\ the eventwise number of reconstructed charged particles as a function of the hard scale $Q\equiv E_{\rm CMS}$;
(ii) the scaled-momentum spectrum
\begin{equation}
x_p \;=\; \frac{2|\mathbf{p}|}{Q}\,, \qquad
\xi_p \;\equiv\; \ln\!\frac{1}{x_p}\, ,
\end{equation}
where $\xi_p$ exposes the hump-backed plateau generated by coherent soft-gluon radiation and local parton-hadron duality;
(iii) the out-of-plane transverse momentum with respect to the thrust system,
\begin{equation}
\pt^{\rm out} \;=\; \big|\mathbf{p}\cdot \hat{\mathbf{n}}_{\rm min}\big|\,,\qquad
\hat{\mathbf{n}}_{\rm min} \;=\; \hat{\mathbf{n}}_T \times \hat{\mathbf{n}}_{\rm maj}\, ,
\end{equation}
with $\hat{\mathbf{n}}_T$ the thrust axis and $\hat{\mathbf{n}}_{\rm maj}$ the major axis that maximises the momentum flow in the plane orthogonal to $\hat{\mathbf{n}}_T$; $\pt^{\rm out}$ therefore measures radiation out of the event plane;
(iv)~rapidities with respect to (a) the thrust axis $\hat{\mathbf{n}}_T$ and (b) the sphericity axis $\hat{\mathbf{n}}_S$,
\begin{equation}
y_T \;=\; \tfrac12\ln\!\frac{E + \mathbf{p}\!\cdot\!\hat{\mathbf{n}}_T}{E - \mathbf{p}\!\cdot\!\hat{\mathbf{n}}_T}\,,\qquad
y_S \;=\; \tfrac12\ln\!\frac{E + \mathbf{p}\!\cdot\!\hat{\mathbf{n}}_S}{E - \mathbf{p}\!\cdot\!\hat{\mathbf{n}}_S}\, ,
\end{equation}
where $\hat{\mathbf{n}}_S$ is the principal eigenvector of the normalised momentum tensor
\begin{equation}
S^{\alpha\beta}\propto \sum_i p_i^\alpha p_i^\beta \Big/ \sum_i |\mathbf{p}_i|^2\, .
\end{equation}
These observables jointly constrain the cluster mass spectrum and fission partition through the low- to mid-$x_p$ and $\xi_p$ shapes, and residual soft wide-angle activity through $\pt^{\rm out}$ and~$y_{T,S}$. The flavour and diquark weights are varied as part of the hadronization model; the present fit includes identified-hadron rates, among them baryons and strange hadrons, but not their momentum spectra.

Taken together, the LEP charged-particle benchmarks in figure~\ref{fig:other-obs} show a clear, systematic improvement over the default Herwig~7.3 settings. The energy dependence of the charged multiplicity $N_{\rm ch}(Q)$ is tightened across the LEP scan, with both normalisation and slope brought into closer accord with the data. In the single-particle spectra, the tuned prediction tracks the $x_p$ fall-off and the $\xi_p=\ln(1/x_p)$ hump-backed plateau more faithfully, indicative of a better-shaped cluster mass spectrum and fission partition. The out-of-plane transverse momentum $\pt^{\rm out}$ shows a softer tail after tuning, consistent with a modest reduction in residual wide-angle activity without degrading the very-low-$\pt^{\rm out}$ core. The rapidity densities $y_T$ and $y_S$ with respect to the thrust and sphericity axes likewise show flatter central plateaux and improved edge fall-offs, reflecting a better description of inclusive charged-particle production within the cluster decay model. Importantly, these gains are uniform across centre-of-mass energies, signalling a more universal description of hadronization and a perturbative shower that is consistent with the diagnostics presented earlier.

Figure~\ref{fig:more-obs} shows a further validation against jet and event shape data. The jet shapes from CDF~\cite{CDF:2005prv}, with a lower centre-of-mass energy and a different mix of quark and gluon jets, are described as well as those from the LHC. The jet rates from ALEPH~\cite{ALEPH:2003obs}, even up to 5 jets, and over a broad range of energies, are significantly improved, as are the flavour-selected (non-$b$ event) event shapes from L3~\cite{L3:2004cdh}.

In short, the tune improves Herwig's predictive power for both hadronization-dominated and IRC-safe $e^+e^-$ observables, while avoiding over-corrections in any single region of phase space.

\newpage


\section{Concluding Remarks and Outlook}
\label{sec:conc}

We have tested the impact of interleaving QCD, QED and EW emissions in the angular-ordered parton shower of Herwig~7. By scanning three interleaving choices ($\qcdonly$, $\qcdqed$, $\qcdqedew$) together with three AO-compatible kinematic reconstruction schemes ($\pt$-preserving, virtuality-preserving, and dot-product-preserving) against QCD-sensitive benchmarks at LEP, Tevatron and the LHC, we find that the interleaving-induced effect is smaller than the scheme-induced envelope for the observables considered here, as summarised in table~\ref{tab:envelopes}. We then tuned the $\dotqcdqedew$ shower and cluster hadronization model parameters, which reduces $\chi^2_{\rm red}$ by about $36\%$ relative to the Herwig~7.3 release values and shows improved LEP charged-particle spectra and multiplicities and consistent LHC jet-substructure observables compared to before, without degrading the Sudakov regions in thrust, $C$, broadening, Durham $y_{23}$, jet mass, width, angularities, or jet shapes.

We hope to use our analysis as a guide to propose the $\dotqcdqedew$ setting as the default choice, as opposed to the current $\qcdqed$, for future releases of Herwig~7. We find this to be a reasonable choice, as not only does it cover a larger set of possible emissions that can occur in a high-energy collision, but it also allows users to use Herwig to study QED/EW radiation in hadron collisions, without worrying about losing the quality of modelling in the QCD-sensitive observables, as studied here. In addition, the simultaneous fit to leptonic and hadronic observables returns compatible ISR and FSR couplings. We plan to take lessons from this study to improve tuning for upcoming releases, where we expect new hadronization and shower features to be available.

\section*{Acknowledgments}

We would like to express our gratitude to our fellow Herwig authors for their invaluable
contributions, with special thanks to Peter Richardson for insightful discussions and support and James Whitehead for comments on the manuscript.
This work has received funding from the European Union's Horizon 2020 research
and innovation programme as part of the Marie Sk\l{}odowska-Curie Innovative Training
Network MCnetITN3 (grant agreement no.~722104). MRM acknowledges support from the
UK Science and Technology Facilities Council (STFC) through the IPPP consolidated grant
ST/X003167/1. SS would like to thank
STFC for the award of a studentship. MHS acknowledges STFC support through grant
ST/X00077X/1.

\bibliographystyle{JHEP}
\bibliography{references}

\end{document}